\documentclass[conference,compsoc]{IEEEtran}
\usepackage{cite}

\usepackage{subcaption}
\usepackage{csquotes}
\usepackage{tabularx}
\usepackage{booktabs}
\usepackage{comment}
\usepackage{multirow}
\usepackage{amsmath}
\usepackage{soul}
\usepackage{xspace}
\usepackage{xcolor,colortbl}

\definecolor{Gray}{gray}{0.85}
\definecolor{LightCyan}{rgb}{0.88,1,1}
\usepackage{stfloats}
\usepackage{bbm} 

\newcolumntype{a}{>{\columncolor{Gray}}r}

\newcolumntype{b}{X}
\newcolumntype{s}{>{\hsize=0.30\hsize}X}

\ifCLASSINFOpdf
\else
\fi

\usepackage{cite}
\usepackage{graphicx}
\usepackage{array}
\usepackage{multirow}
\usepackage{pifont}
\usepackage{color, colortbl}
\usepackage{siunitx}
\usepackage{booktabs}
\usepackage{rotating}
\usepackage{xparse}
\usepackage{tabularx}
\usepackage{url}
\usepackage{pifont}
\usepackage[font=small]{caption}
\usepackage{subcaption}
\usepackage{tikz}
\usepackage{makecell}
\usepackage{xspace}

\newcommand*{\escapeI}[1]{\texttt{\expandafter\string\csname #1\endcsname}}

\begin{document}
%

\title{Specious Sites: Tracking the Spread and Sway of Spurious News Stories at Scale}
%
%
%
\author{\IEEEauthorblockN{Hans~W.~A.~Hanley}
\IEEEauthorblockA{
hhanley@cs.stanford.edu \\
\textit{Stanford University}}
\and
\IEEEauthorblockN{Deepak Kumar}
\IEEEauthorblockA{
kumarde@cs.stanford.edu \\
\textit{Stanford University}}
\and
\IEEEauthorblockN{Zakir Durumeric}
\IEEEauthorblockA{
zakir@cs.stanford.edu \\
\textit{Stanford University}}
}
\markboth{Journal of \LaTeX\ Class Files,~Vol.~14, No.~8, August~2015}%
{Shell \MakeLowercase{\textit{et al.}}: Bare Demo of IEEEtran.cls for IEEE Journals}

\maketitle

\begin{abstract}
Misinformation, propaganda, and outright lies proliferate on the web, with some narratives having dangerous real-world consequences on public health, elections, and individual safety. However, despite the impact of misinformation, the research community largely lacks automated and programmatic approaches for tracking news narratives across online platforms. In this work, utilizing daily scrapes of 1,334~unreliable news websites, the large-language model MPNet, and DP-Means clustering, we introduce a system to automatically identify and track the narratives spread within online ecosystems. Identifying 52,036~narratives on these 1,334~websites, we describe the most prevalent narratives spread in 2022 and identify the most influential websites that originate and amplify narratives. Finally, we show how our system can be utilized to detect new narratives originating from unreliable news websites and to aid fact-checkers in more quickly addressing misinformation. We release code and data at \url{https://github.com/hanshanley/specious-sites}. 
\end{abstract}


%
\IEEEpeerreviewmaketitle

\section{Introduction}
Over the last decade, spurious, misleading, and outright false information has spread throughout online ecosystems~\cite{Stanton2020}. Digital misinformation has influenced elections~\cite{bovet2019influence}, promoted bogus health cures leading to unnecessary deaths~\cite{ball2020epic}, and incited mob violence throughout the world~\cite{banaji2019whatsapp,hanley2022no}. Worsening the problem, misleading stories have been shown to spread at over ten times the rate of true information~\cite{vosoughi2018spread}. 

The security community, likening disinformation to an attack similar to spam, phishing, and censorship~\cite{thomas2021sok, zurko2022disinformation}, has applied a range of methodologies to ameliorate its spread~\cite{rajdev2015fake,wu2019misinformation,saeed2022trollmagnifier,hounsel2020identifying,kaiser2021adapting,paudel2023lambretta}. For example, by examining features similar to those used to identify spam accounts, researchers have identified networks of state-propagandists throughout Reddit and Twitter~\cite{saeed2022trollmagnifier, zannettou2019disinformation}. However, despite these advances, most investigations into false narratives remain limited in scope and retroactive, primarily conducted through time-consuming, qualitative approaches~\cite{plasser2005hard, leskovec2009meme}. To make fundamental progress in combating the threat posed by disinformation, we argue that the security community must build approaches for tracking the spread of false narratives globally and in real time.

In this work, we present an NLP-based approach for programmatically identifying and tracking the spread of {narratives and stories} across \emph{unreliable} news websites and social media platforms. Between January 1 and November 1, 2022, we crawl a set of 1,334~known misinformation, state-propaganda, biased, and otherwise {unreliable} news websites as well as two fringe forums, 8kun and 4chan. We extract passages from these news articles, which we embed using an MPNet model~\cite{song2020mpnet} fine-tuned with contrastive learning on the semantic textual similarity task. Employing a modified version of the nonparametric algorithm DP-Means, we cluster these embeddings to identify specific {narratives/stories}.

Our approach enables us to isolate and track 52,036~narrative threads that spread on {unreliable} news websites during 2022. We do not attempt to determine whether individual stories are factual, which is a qualitative task that ML-based approaches have failed to reliably achieve. Rather, we track all narratives from these {specious sites} across online ecosystems and quantify these websites' influence. We find, unsurprisingly, that many of the most prominent narratives in 2022 concerned the Russo-Ukrainian War and inflation, with websites like {southfront.org}, {rt.com}, and {zerohedge.com} dominating these topics. Identifying which websites play outsized roles in \emph{originating} and \emph{amplifying} narratives across our set of {{unreliable}} news websites, we find that a website's popularity has a small correlation with its ability to propagate narratives with seemingly minor websites like {infostormer.com} or {barenakedislam.com} playing massive roles in popularizing stories. 

Next, we investigate how our method can be used to focus limited investigative resources on the most pernicious narratives. We show that, like an early detection alarm, our approach can identify when new narratives emerge. Comparing when popular false narratives appeared to when three major organizations (AP News, Reuters, and Politifact) fact-checked them, we show that our system can prioritize checking misleading narratives {months} before their peak when they first start to gain traction. We hope that this type of real-time visibility can enable fact-checkers and journalists to more efficiently track and respond to new, potentially problematic narratives as soon as they emerge.

Similar to past large-scale empirical analysis in the security community (\textit{e.g.},~\cite{durumeric2013zmap, moore2003inside, meiklejohn2013fistful, kanich2008spamalytics, motoyama2011analysis, antonakakis2017understanding, mccoy2012pharmaleaks, afroz2014doppelganger, zeng2020bad, heninger2012mining}), our work shows that a programmatic approach to tracking narratives at scale uncovers a set of online propagation patterns that would have been difficult to uncover through manual, small-scale investigations. We also discuss how having a {continuous} tracking process can help analysts uncover and track the most worrisome influence operations. We stress that our approach does not make factual judgments on individual stories or on the reliability of websites. Indeed, our system takes \textit{as input} websites that human experts previously labeled as {unreliable}. Rather, our approach provides the critical, real-time visibility into the spread of news narratives online that human experts need to effectively identify and respond to misinformation.

%
%
%
%



\section{Background\label{sec:background}}

Unreliable information often spreads through multiple avenues as individual users, state-supported actors, and even entire platforms participate in the dissemination of falsities. Unreliable information can take the form of \textit{misinformation}, \textit{disinformation}, \textit{fake-news}, \textit{propaganda}, among others~\cite{jack2017lexicon}. \textit{Misinformation} is any information that is false or inaccurate regardless of the author's intent~\cite{jiang2018linguistic,lewandowsky2012misinformation,jack2017lexicon,allcott2019trends}. The term ``fake news'' is often used interchangeably with misinformation. Disinformation, in contrast, is inaccurate information spread with the deliberate purpose to mislead~\cite{akbar2021misinformation,jack2017lexicon}. Similar to disinformation, \textit{propaganda} refers to ``deliberate, systematic information campaigns, usually conducted through mass media forms'' regardless of whether the information is true or false~\cite{jack2017lexicon}. A single narrative can be considered \textit{disinformation} when spread by a state actor and as \textit{misinformation} when spread by users. Individual websites can have mixtures of misinformation, disinformation, propaganda, and true information~\cite{starbird2018ecosystem}. We refer readers to Jack et~al.~\cite{jack2017lexicon} for a more detailed taxonomy.

\begin{figure*}
  \centering
  \includegraphics[width=0.79\linewidth]{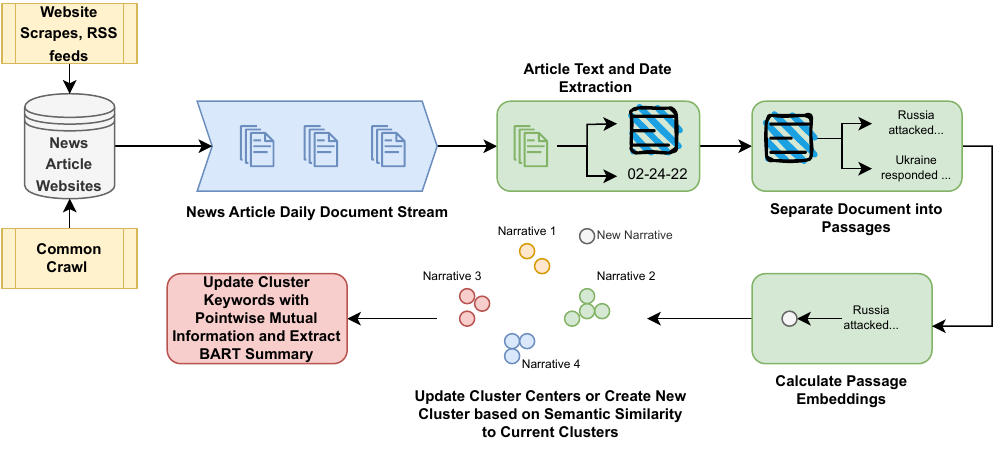}

\vspace{1pt}
\caption{Our pipeline for identifying and labeling narrative clusters from the daily publications of {unreliable} news websites.}
\label{fig:pipeline}
\vspace{-15pt}
\end{figure*}

\section{Methodology}
The goal of our work is to programmatically track how narratives spread amongst ``\textit{unreliable}'' news websites. In this section, we define a narrative and then describe how we collect news articles and extract narratives. We emphasize that while we focus on websites known to publish misleading, false, or state-controlled narratives, we do not assume that {all} narratives from these sites are ``misinformation''. Indeed, many stories are not~\cite{starbird2018ecosystem}. We do not label new narratives as ``misinformation'', which is a qualitative, investigative task. As such, we refer to only {individual} stories that have been previously identified by experts as \textit{misinformation} or \textit{disinformation}, as false. We label the websites that spread these verified false narratives as \textit{unreliable}. 


\subsection{Narrative Definition}

Tracking misinformation narratives requires a high degree of specificity. Unlike traditional topic modeling, which seeks to identify \textit{themes} and/or statistical word correlations~\cite{allan2002detection,devine2022unsupervised,jelodar2019latent}, misinformation tracking requires distinguishing between specific narratives and stories. Within this work, we define a narrative/story using the same definition as the Event Registry~\cite{leban2014event}, Hanley et~al.~\cite{hanley2022happenstance}, and Miranda et~al.~\cite{miranda-etal-2018-multilingual}: collections of documents that seek to address the same \textit{event} or \textit{issue}. For example, two example events in the Event Registry are ``Felix Baumgartner’s jump from a helium balloon on October 14, 2012'' and ``bombings during the Boston Marathon on April 15, 2013.'' Within our dataset, events constitute ideas like ``election fraud in the 2020 US election'' and ``the COVID-19 vaccine leading to mass death.'' An example of two ideas---while related---that we do not consider to be the same narrative are ``US funds Ukrainian War'' and ``Russia attacks Ukraine.''

\subsection{System Architecture}

As shown in Figure~\ref{fig:pipeline}, our system (1) collects news articles from {{unreliable}} news sites through web scraping on a daily basis and from Common Crawl data~\cite{CommonCrawl2022}; (2) parses out the news articles, extracting article text and segmenting articles into constituent passages; and (3) embeds text passages into a shared subspace utilizing the large language model MPNet~\cite{song2020mpnet}. To track the spread of narratives, our system then (4) clusters semantically similar content using DP-means, and (5) extracts keywords and generates summaries of the clusters. 

We opt for this LLM-based approach because our system needs to track specific narratives. Prior approaches that utilize simpler, more generic keyword-based topic modeling tools like LDA fall short in identifying specific narratives across different news websites~\cite{min2015cross}. Furthermore, keyword-based approaches often rely on pre-existing expert knowledge of disinformation campaigns and largely cannot adapt to the rapid pace of the news ecosystem~\cite{bal2020analysing}. By utilizing this new approach, our system can update and track news stories without \textit{a priori} or domain-specific knowledge in an efficient and fine-grained manner.

For this paper, we use data from January 1 to November 1, 2022, but we emphasize that our system runs continuously, enabling us to identify new narratives in near real time. In the rest of this section, we detail each step of our methodology and validate that our system captures specific and coherent narratives.

\subsection{Data Collection}
Our study is based on scraping and parsing articles from websites known to spread unreliable information. 

\vspace{2pt}
\noindent
\textbf{Unreliable News Websites.}
We collect articles from 2,514~candidate websites that have been labeled as ``politically biased'', ``misinformation'', ``disinformation'', ``conspiracy'', ``fake news'', or ``state-based propaganda'' by past studies (Iffy Index~\cite{iffy2022},
OpenSources~\cite{fakenews2020},
Politifact~\cite{politifact2017}, Snopes~\cite{fake2017},
Melissa Zimdars~\cite{melissa027},
and Hanley et~al.~\cite{hanley2022golden}). This list includes politically-biased websites like {dailywire.com},
conspiracy-oriented websites like {x22report.com}, and state-propaganda outlets like {rt.com}. 

\vspace{3pt}
\noindent
\textbf{Scraping Articles.}
We crawl websites using Colly\footnote{https://github.com/gocolly/colly} and Headless Chrome orchestrated with Python Selenium. For each website, we collect the homepage and linked pages daily from January 1 to November 1, 2022. To ensure full coverage of each site's published articles, we additionally gather the HTML pages indexed by Common Crawl~\cite{smith2013dirt} 
for each site during this same period. We emphasize that under 1\% of articles were found only in the Common Crawl dataset, indicating that our scrapes found the vast majority of published content on each site. We then parse each HTML page to collect the published \emph{articles} using the Python libraries \texttt{newspaper3k} and \texttt{htmldate}. There were several instances (\textit{e.g.}, {sputniknews.com}) where this approach failed; in these cases, we built custom parsers based on site-specific HTML elements.

Of our 2,514~candidate websites, 1,334 were operational and published articles during our 2022 measurement period (many sites that spread {unreliable} information are short-lived~\cite{hounsel2020identifying, dahlke2023pie}). Altogether, we collect 1,915,449~articles. We provide the URL data to researchers upon request.

\subsection{Preprocessing and Embedding \label{sec:preprocess}}
To prepare data for embedding, we first remove any non-English articles using the Python \texttt{langdetect} library and then remove URLs, emojis, and HTML tags. We then segment each article into its constituent paragraphs by splitting article text based on newline and tab characters. Then, in line with prior work, we subsequently divide these paragraphs into  100-word article \emph{passages}~\cite{piktus2021web, hanley2022happenstance, hanley2022partial}. 

\begin{table}
    \centering
    \setlength\tabcolsep{4pt}
    \small
    \begin{tabular}{ccccc}
    {GloVe}  & {BERT}  & {USE} &{All-MPNet} &{Our Model} \\ \toprule
    0.580 & 0.464 & 0.749 &0.840 &\textbf{0.856} \\
    \end{tabular}
    
    \caption{ Evaluation--based on Pearson Correlation---of our MPNet-contrastive model and other models on the SemEval STS-benchmark~\cite{cer2017semeval}. Data for GloVe~\cite{pennington2014glove}, BERT~\cite{devlin2019bert}, the Universal Sentence Encoder (USE)~\cite{cer2018universal} is from Reimers et~al.~\cite{reimers2019sentence}.}
    \label{tab:sts-benchmark}
    \vspace{-15pt}
\end{table}

After preprocessing, we embed the constituent passages that make up each article. Embedding passages rather than entire articles is in line with prior work~\cite{piktus2021web} for topic analysis as articles often address multiple narratives but embeddings should represent only a single narrative or idea~\cite{piktus2021web, hanley2022happenstance,hanley2022partial}. We thus embed passages to capture context while also obtaining an embedding for the (often) one narrative/idea present within the passage.

We specifically embed passages using a version of MPNet that we fine-tune on the semantic text similarity (STS) task~\cite{rondinelli2022zero, hanley2022happenstance} using unsupervised contrastive learning for sentence embeddings as specified in Gao et~al.~\cite{gao2021simcse} on a random assortment of passages from January 2022 from our websites. We perform this fine-tuning with the default hyperparameters (learning rate $3\times 10^{-5}$, batch size=128, and 1M examples) specified in Gao et~al. and by freezing all but the last two layers of a public version of MPNet.\footnote{\url{https://huggingface.co/sentence-transformers/all-mpnet-base-v2}} See Appendix~\ref{sec:contrastive} for details. This ensures that our model is attuned to the language present on our set of websites. As seen in Table~\ref{tab:sts-benchmark}, despite not being trained on the SemEval STS Benchmark~\cite{cer2017semeval}, a benchmark for measuring the quality of text embeddings, our model outperforms the fine-tuned publicly released version of MPNet. After fine-tuning our model, from the 1.9M~articles, we embed 25,337,614
~passages (10~hours on an Nvidia RTX A6000).

\begin{figure}
  \centering
  \includegraphics[width=1\columnwidth]{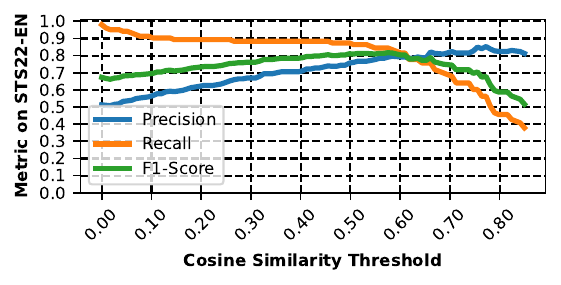}
\caption{Evaluation of our model's precision, recall, and $F_1$ scores on the English portion of the SemEval22 dataset~\cite{goel2022semeval} (using 3.0 as the cut-off for the two articles being about the same event~\cite{hanley2022partial}).
}
\label{fig:semeval-study}
\vspace{-15pt}
\end{figure}

\subsection{Comparing Semantic Content\label{sec:comparing} }
We compare the semantic content of our embeddings utilizing cosine similarity~\cite{song2020mpnet}. Prior work~\cite{hanley2022happenstance,vetter2022using,bernard2022tracking} has found that a cosine similarity threshold of 0.60--0.80 can be used to determine whether two pieces of text are about the same topic. However, to ensure that we select a minimum threshold that accurately models whether passages are about the same \textit{narrative} as defined in Section~\ref{sec:background}, we: (1) benchmark our model on the English portion of the Multilingual SemEval2022 dataset~\cite{goel2022semeval}, and (2) manually validate the coherency of a random sample of passage pairs.
\begin{table}
    \centering
    \setlength\tabcolsep{3pt}
    \footnotesize
    \begin{tabular}{cccc}
    {0.45  Thresh.}  & {0.50 Thresh.}  & {0.55 Thresh.} &{0.60 Thresh.} \\ \toprule
    $80.6\% \pm 6.33\% $  & $85.7\% \pm 8.29\% $ & $94.7\% \pm 6.94\%$&$   96.29\% \pm 7.12\%$ \\ 
    \end{tabular}
    \caption{ Evaluation of the precision of embedded messages having the same narrative at various thresholds utilizing 2000~random passages. We provide 95\% Normal confidence intervals.}
    \label{tab:similar-topic-precision}
    \vspace{-5pt}

\end{table}
\begin{figure}[]
\noindent\fcolorbox{black}{lightgray}{%

\begin{minipage}{.47\textwidth}
\tiny
\textbf{PASSAGE 1:} The MaraLago search warrant served Monday was part of an ongoing Justice Department investigation into the discovery of classified White House records recovered from Trump's home earlier this year. The Archives had asked the department to investigate after saying 15 boxes of records it retrieved from the estate included classified records.

\textbf{PASSAGE 2:} The FBI raided Donald Trump's estate in MaraLago using the pretext of Trump supposedly violating the Presidential Records Act by keeping documents after he left office to initiate a siege to terrorize Joe Biden's chief political rival.
\end{minipage}}
\caption{Passage pair at our selected similarity threshold (0.60).}
\label{figure:paragraph_pairs}
\vspace{-10pt}
\end{figure}

As seen in Figure~\ref{fig:semeval-study}, on the SemEval22~\cite{goel2022semeval} dataset, as the similarity threshold increases, our model's precision in determining whether two passages are about the same narrative increases while the corresponding recall decreases, reaching a peak $F_1$ score near 0.60 cosine similarity. 

To confirm this result, we perform a manual evaluation by selecting 2,000~random passage pairs from our dataset with similarities at varying thresholds and have two experts determine whether the passage pairs are about the same narrative (per our definition), determining the corresponding precision at various thresholds. We calculate a Cohen's Kappa of 0.80 between our two raters, indicating a high degree of agreement. As seen in Table~\ref{tab:similar-topic-precision}, as the threshold used to determine similarity increases, we see an increase in precision. We choose a threshold of 0.60 (as a lower bound) as it has acceptable manually calculated precision (96.29\%) and $F_1$ score on the SemEval22~\cite{goel2022semeval} dataset (0.809). We present an example passage pair at our selected threshold of 0.60 in Figure~\ref{figure:paragraph_pairs} and at other thresholds in Appendix~\ref{sec:thresholds}.

\begin{table*}
\centering
\scriptsize
\begin{minipage}{.46\textwidth}
\setlength{\tabcolsep}{5pt}
\centering
\begin{tabular}{llrr}
\toprule
 &   &  Passages & \\
Narr. &{Keywords} & Checked & Prec. \\\midrule
1 & kiril, patriarch, orthodox, church, putin&427 &99.53\% \\
2 & abbott, texas, border, greg, lone& 500 & 99.40\% \\
3 & sinema, manchin, filibuster, kyrsten, senate& 500 & 100.00\%\\
4 & hurricane, atlantic, storm, tropic, season & 500 & 99.80\% \\
5 &balloon, leaflet, korea, korean, north & 100 & 100.00\%\\
6 &monkeypox, york, nyc, outbreak, city & 385 & 96.62\%\\
7 &johnson, resign, poll, boris, tory & 500 & 95.20\%\\
8 &antibody, monoclon, regeneron, omicron, variant & 302 & 100.00\%\\
9 &fauci, anthoni, kennedy, pharma, gate & 384 & 100.00\% \\
10 &nucleic, acid, test, shanghai, province & 500 & 99.40\%\\
11&finland, sweden, nato, deploy, nuclear & 133 &100.00\%\\
12&windfall, profit, tax, oil, barrell& 379 & 99.21\%\\
13 & energy, europe, crisis, price, electric & 500 & 100.0\% \\
14&pen, macron, le, french, marin & 500 & 99.20\% \\
15& protein, spike, mrna, inject, cell & 173 & 100.00\%\\
\bottomrule
\\
\end{tabular}
\end{minipage}
\begin{minipage}{.52\textwidth}
\centering
\setlength{\tabcolsep}{5pt}
\begin{tabular}{llrr}
\toprule
 &   & Passages  & \\
Narr. &{Keywords} & Checked & Prec. \\\midrule
16 &peterson, suspend, rubin, jordan, elliot &252 & 100.00\% \\
17& norwegian, ellingsen, feminist, lesbian, norway & 108 &  100.00\% \\
18& lantsman, trudeau, melissa, mp, swastika& 127 & 100.00\% \\
19& polio, 1979, eradicate, virus, disease& 109 & 99.08\%\\
20&humanitarian, aid, shelter, refuge, relief & 500 & 100.00\%\\
21& kiev, coup, nationalist, neonazi, nazi & 143 & 98.60\% \\
22 & noah, flood, ark, wives, genesis & 155 & 90.97\%\\
23 &fda, prescribe, offlabel, drug, treatment& 195 & 100.00\%\\
24 & refugee, asylum, persecute, seeker, migrant& 192 &100.00\%\\
25& swift, sanction, bank, sberbank, vtb& 500& 100.00\%\\
26& orban, fidesz, viktor, hungary,victory& 469 & 99.60\%\\
27&unvaccined, infection, recipe, covid19, covid&  167 & 100.00\% \\
28&nuclear, closer, brink, cuban, war & 364 & 96.43\% \\
29&alexandra, pelosi, footage, nancy, daughter& 100 & 100.00\%\\
30& civilian, kabul, afghan, drone, strike & 500 &94.00\% \\
\bottomrule
 && \textbf{Prec.}& \textbf{98.90\%}
\end{tabular}
\end{minipage}
\caption{ Evaluation of the precision of our narrative analysis model on a random set of 30~stories/narratives derived from the articles in our dataset. Keywords were extracted utilizing pointwise mutual information. We checked all available passages in cases where there are fewer than 500 passages in the story/narrative cluster. } 
\label{tab:precision-test}
\vspace{-17pt}
\end{table*}

\subsection{Identifying Narratives\label{sec:idnarratives}}
To identify \textit{narrative/stories} in our dataset, we cluster our passage embeddings using cosine similarity and DP-Means, a non-parametric version of K-means (Appendix~\ref{sec:dpmeans}). Prior work has identified narratives using a similar high-level approach~\cite{leskovec2009meme,hanley2022happenstance}, but our methodology differs in several ways based on our unique requirements. First, our approach must be \emph{highly scalable}. While Hanley et~al.~\cite{hanley2022happenstance} utilize a BERTopic-based method~\cite{grootendorst2022bertopic}, we find that this does not scale to the approximately 100K~embeddings per day in our dataset. Second, we need to update our clusters on a daily basis as new news articles are published, which past approaches like BERTopic do not allow.  Third, since the number of narratives is unknown \textit{a priori}, the methodology must automatically infer the number of clusters, which precludes parametric algorithms like incremental K-means clustering.

We specifically adapt Dinari et~al.'s efficient and parallelizable version of DP-Means~\cite{dinari2022revisiting} (Appendix~\ref{sec:dpmeans}), making four alterations. First, we cluster embeddings based on their cosine similarity rather than their Euclidean distance. We set $\lambda=0.60$ (the minimum cosine distance an embedding can be from a cluster before a new cluster is created) to ensure that clusters have high semantic similarity, informed by our prior manual investigation (Table~\ref{tab:similar-topic-precision}). Second, we perform partial fits over each day's worth of news article embeddings. Specifically, on each day throughout 2022, we embed that day's passages and update the previous day's cluster centers (\textit{i.e.,} we update our given clusters on a daily basis with the DP-Means algorithm until convergence utilizing that day's article embeddings). Third, we remove the random reinitialization of clusters added by Dinari et~al.~\cite{dinari2022revisiting} from the algorithm; we find that this step often led to over-clustering given that many website passages are slight variations of each other. Lastly, we note that rather than relying on Dinari et~al.'s released code, we re-implement their algorithm to take advantage of the matrix multiplication speedups that come from utilizing a GPU (3~times speedup with an Nvidia RTX A6000). 

For this work, we utilize the clusters from November~1, 2022. From January~1 to November~1, clustering all embeddings required the equivalent of 1.5~days. We filter out clusters where 50\% or more of the passages are from only one website (\textit{e.g.}, author bios) or there were fewer than 25~articles to remove spam, similar to the methodology specified by Leskovec et~al.~\cite{leskovec2009meme}. After this removal, we identify 52,036~narrative clusters. Each article's passages are part of an average of 5.12~narrative clusters (4.0~median). On average, each embedding has an average similarity of 0.688 to its cluster center, which shows that our embeddings are assigned to clusters with high semantic similarity. Each narrative cluster has an average cosine similarity of 0.016 with other identified narrative clusters, which indicates that our approach identified distinct narratives.

\subsection{Interpretability and Narrative Specificity}
We create human-interpretable identities for our narrative clusters using two approaches. First, we extract the most distinctive and representative keywords of the cluster using pointwise mutual information~\cite{hanley2022special,bouma2009normalized} (Appendix~\ref{sec:pmi}). Pointwise mutual information (PMI) is an information-theoretic measure for discovering the associations amongst different entities ~\cite{bouma2009normalized}. As in Kessler et~al.~\cite{kessler2017scattertext}, rather than finding the pointwise mutual information between different words, we utilize the measure to understand individual words' association with narrative clusters. In this manner, we find the set of words most \emph{distinctive} to/associated with each cluster. Second, after identifying the top five passages closest (\textit{i.e.}, with the largest cosine similarity) to the center of the cluster, we use an off-the-shelf state-of-the-art BART~\cite{lewis2020bart} summarization tool from Huggingface fine-tuned on news data to summarize the cluster. We utilize this approach because while keywords provide an identifiable ``handle'' for each cluster, keywords typically do not fully capture the full semantic meaning or the specificity of our narrative clusters. For example, the auto-generated summary for the cluster with keywords \textit{Age, Pfizer, Booster, Children, Vaccine} is:
\begin{displayquote}
\small
\textit{The U.S. Food and Drug Administration FDA in October 2021 authorized the PfizerBioNTech COVID vaccine for children 5 through 11. Children under 5 remain the only segment of the US population that isn't eligible for one of the COVID vaccines.}
\end{displayquote}

\noindent where a random passage from the cluster states: 
\begin{displayquote}
\small
\textit{As of now, U.S. children aged five and older are eligible for the COVID19 vaccine, though only Pfizer's shot has received authorization. The Pfizer jab is also available as a booster for children 12 and older.}
\end{displayquote}
\noindent 
However, the auto-generated summary for a similar cluster with keywords \textit{Children, Risk, Adult, Covid, Immunity} is:

\begin{displayquote}
\small
\textit{Children have a minuscule risk of COVID mortality. There is very limited safety data for vaccines from the trials on children. If the risk of adverse reactions is the same as for adults, the harms outweigh the risks.}
\end{displayquote}
\noindent where a random passage from the cluster states: 
\begin{displayquote}
\small
\textit{COVID poses no danger to children. They have a statistically zero chance of dying from that disease. The COVID shots, however, are already linked to innumerable adverse reactions, and their longterm side effects are unstudied.}
\end{displayquote}

\noindent
This illustrates the need for further specificity using summarization to understand the narratives being spread. We provide several additional examples in Appendix~\ref{sec:cluster-spec}.

\subsection{Validating Narrative Clusters}
\label{sec:verification}

We evaluate our narrative clustering technique by validating whether a random sample of 500~passages (or maximum present) for a random set of 30~narrative clusters are about the same narrative using the methodology outlined in Section~\ref{sec:comparing}.
Our methodology identifies coherent story/narrative clusters with an overall 98.9\% precision and a minimum precision of 90.97\% for Topic 22 (Table~\ref{tab:precision-test}).

\subsection{Ethical Considerations}

Our analysis is based on analyzing publicly posted news articles. We limit the load that each news site experiences by checking for new articles daily at a maximum rate of one request every 10~seconds. 
We further follow the guidelines as outlined by prior work for scraping data~\cite{hanley2022no, durumeric2013zmap}. The hosts that we scan from are identifiable through WHOIS, reverse DNS, and an HTTP landing page explaining how to reach us if they would like to be removed from the study. We received no requests from websites to opt-out.  

\subsection{Positionality Statement}
The misinformation websites we study often covered contentious political issues including election denial, the Russo-Ukrainian War, and US abortion rights. As US-based English-speaking researchers, we inevitably bring some biases to discussing these issues. We attempt to remain as neutral as possible. We do not take any stance on political issues and when labeling specific stories as being misinformation, we rely fully on cited prior work from other researchers and/or news groups.




\begin{table*}
\centering
\scriptsize
\selectfont
\setlength{\tabcolsep}{4pt}
\begin{tabularx}{2.0\columnwidth}{lsrrsb}
\toprule
Narr. &  {Keywords} & Articles &  Websites & Most Profilic Domains &  Auto-Generated Summary\\ \midrule
 1 & ukraine, troop, kyiv, russian, donbas &9,579& 378 &express.co.uk (626), southfront.org (523), dailymail.co.uk (464) & The Russian military has not been able to fully encircle and neutralize the grouping of Kyiv s forces in the Donbass so far. At the same time, the Russians managed to liberate a number of important territories and towns.  \\ \midrule
 2 & zelensky, volodymyr, ukraine, kyiv, president &8,705 & 392 & dailymail.co.uk (548), nypost.com (466), express.co.uk (411) & Ukrainian President Volodymyr Zelensky has accused Russian forces of committing genocide in his country. He also slammed the West.\\ \midrule
 3 & index, consumer, inflation, cpi, price & 7,240 & 444  &shorenewsnetwork.com (963), theepochtimes.com (514), dailymail.co.uk (336) & The consumer price index climbed 0.6 percent from a month before. Compared with January of last year, consumer prices are up 7.5 percent. The Consumer Price Index increased 9.1 percent in the year through June. \\ \midrule

 4 & musk, elon, twitter, platform, tesla & 7,196 &335 & nypost.com (364), dailymail.co.uk (281), theepochtimes.com (222) & Tech Mogul and Tesla Boss Elon Musk is wellknown for his wisecracks and witty posts he shares on Twitter. Musk has been critical of social media, particularly Twitter, over its enforcement of rules that critics say targets conservative voices. \\ \midrule
 5 &  germany, europe, oil, sanction, energy & 6,812 & 362 & express.co.uk (355), zerohedge.com (323), rt.com (283) & Russia has been hit by sweeping sanctions on its economy and trade since the start of Putin's war in Ukraine. But measures by EU governments have not targeted oil and gas contracts with Moscow. Europe is heavily reliant on Russia for its energy needs.\\

\bottomrule
\end{tabularx}
\caption{Top 5 narratives---by number of articles---in our 2022 dataset.} 
\vspace{-10pt}
\label{tab:narratives} 
\end{table*}

\begin{figure*}
  \centering
  \includegraphics[width=\linewidth]{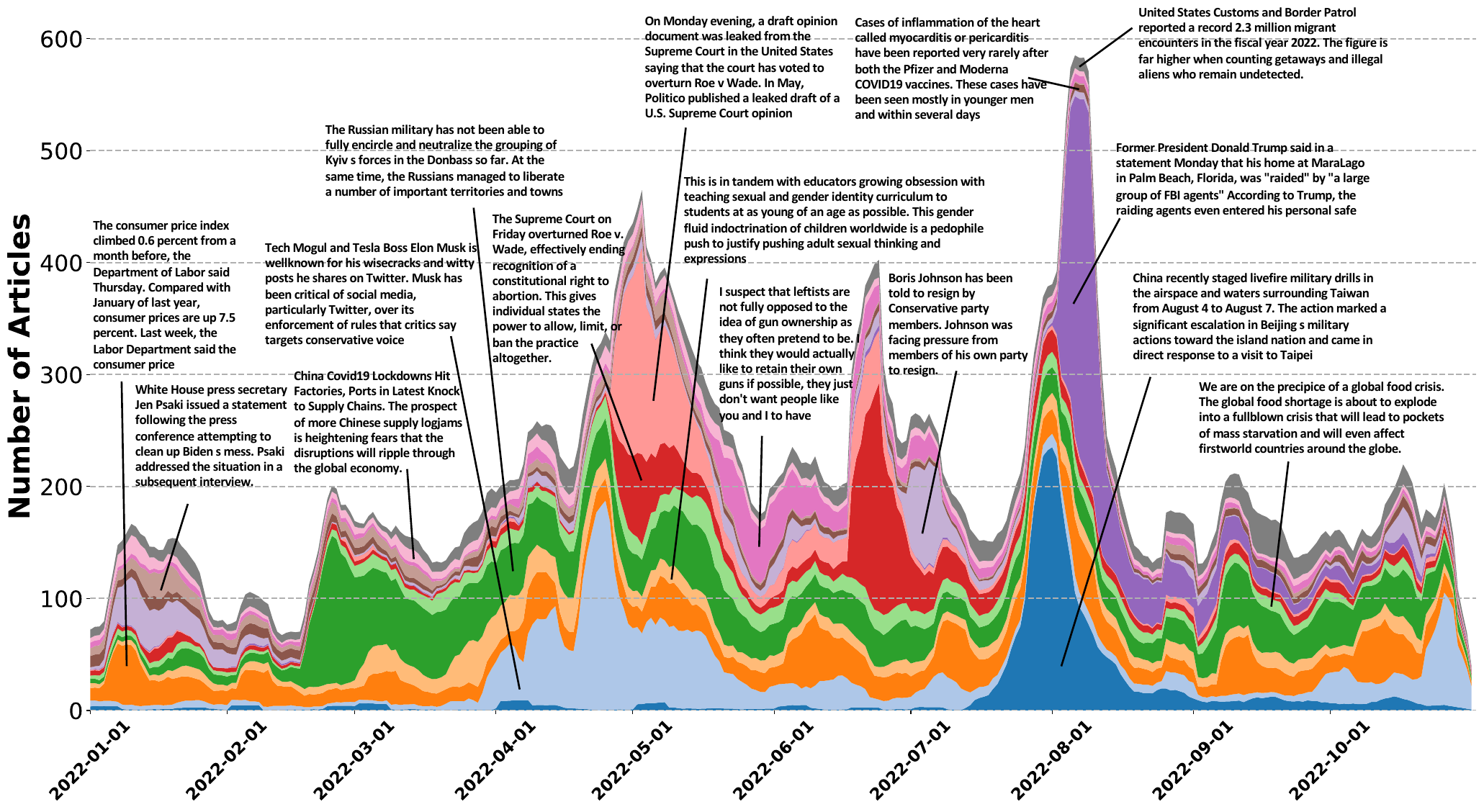}
\caption{Article volume of popular narratives from January 1, 2022, to November 1, 2022.}
\label{fig:website-domains-over-time}
\vspace{-15pt}
\end{figure*}

\section{Narratives on Unreliable News Sites}

In the last section, we presented and validated our methodology for programmatically extracting the {narratives} promoted by {unreliable} news websites. Here, we describe the most prolific narratives, trace three misinformation/propaganda stories, and derive communities of topically-related websites.

\subsection{The Largest Narratives\label{sec:largest}}

We start by analyzing the narratives most prolifically covered by our set of {unreliable} news sites in 2022. As can be seen in Table~\ref{tab:narratives}, the most popular narratives concerned the Russo-Ukrainian War, inflation, and Elon Musk's criticism and later acquisition of the social media platform Twitter. As can be seen in Figure~\ref{fig:website-domains-over-time}, we observe peaks in coverage of specific stories, as well as narratives that maintained consistent coverage throughout our study. For example, stories about abortion peak both before the US Supreme Court decision (Dobbs v.\ Jackson) about federal abortion rights was leaked and following the official decision~\cite{StaffAborrtion2022}. In contrast, a narrative about the EU's role in NATO saw a steady stream of articles throughout the year, with a slight uptick following the Russian invasion of Ukraine. Analyzing the specific news sites that post about each narrative, we find that many Russian-backed and controlled websites~\cite{center2020pillars} such as {rt.com} and {southfront.org}, in addition to several UK-based tabloids {express.co.uk} and {dailymail.co.uk} were the most prolific in writing about the Russian invasion of Ukraine (Narratives~1, 2, 5 in Table~\ref{tab:narratives}). This largely matches previous studies of the Russian-controlled media in influencing discussions on the war~\cite{hanley2022happenstance}.

\subsection{Misinformation/Propaganda Case Studies\label{sec:case-studies}}
As seen in the last section, many of the most common narratives are mainstream news topics. However, one of our goals is to track the spread of misinformation narratives. In this section, we show that our technique is capable of tracking known unreliable narratives by investigating the evolution of one confirmed \textit{propaganda} and two confirmed \textit{misinformation} stories.

\begin{figure}
  \centering
  \includegraphics[width=\columnwidth]{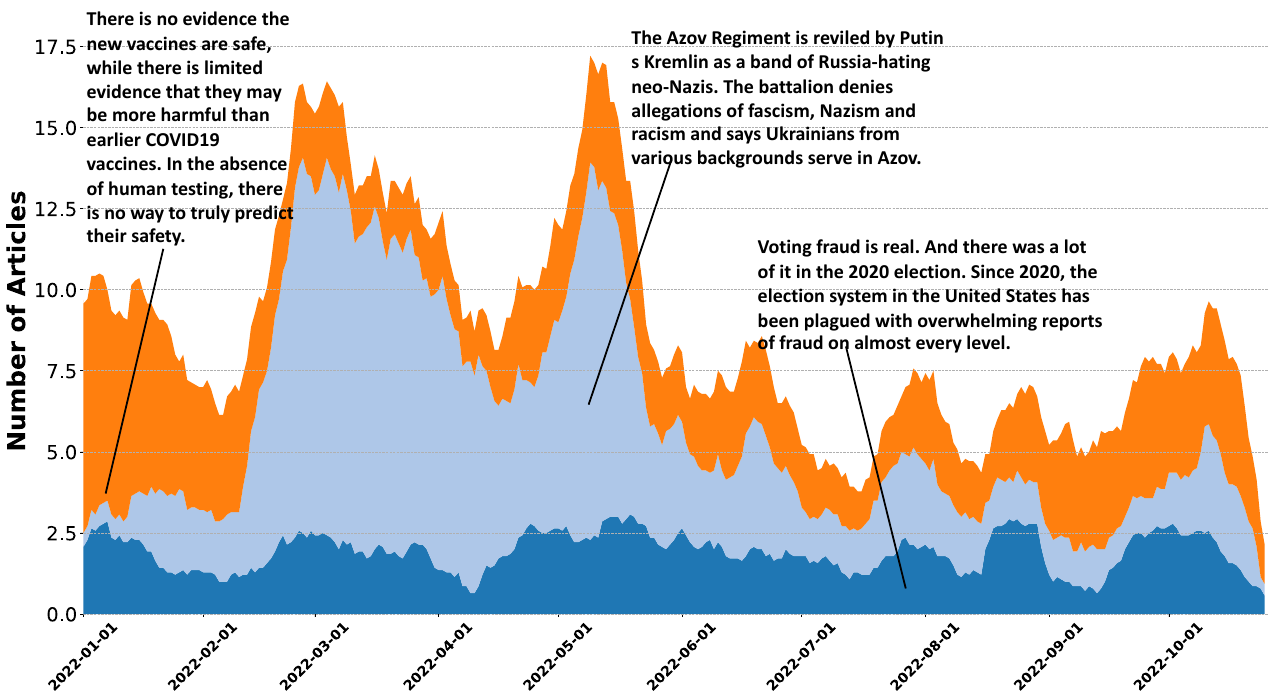}
\caption{ Volume over time for case-study narratives of Ukrainian Nazis, Killer COVID-19 vaccines, and 2020 Election Denialism. 
}
\label{fig:case-study}
\vspace{-15pt}
\end{figure}

\vspace{3pt}
\noindent \textbf{Ukrainian Nazis} (Keywords: Azov, Battalion, Regiment, Far-right, Ukraine):
One of the most prominent propaganda narratives utilized by Russian media in justifying the Russian Federation's invasion of Ukraine was that the Ukrainian government was controlled by ``neo-nazis''~\cite{hanley2022special}. This is despite Ukraine's relatively low level of antisemitism~\cite{Masci2018}. 
Our method is able to find that even before the Russian invasion of Ukraine on February 24, 2022, there were heavy references to Nazism in Ukraine by Russian-controlled or influenced outlets. For example, on January 27, {gloablresearch.ca} penned:\footnote{\url{https://web.archive.org/web/20220127162858/https://www.globalresearch.ca/war-fever-air-west-confuses-russia-nazi-germany/5768335}}
\begin{displayquote}
\small
\textit{If we are to draw parallels between the current crisis on the Ukraine border and WW2 we should compare the Neo-Nazi ideology which dominates Ukrainian nationalism with that of Nazi Germany.} 
\end{displayquote}
However, as seen in Figure~\ref{fig:case-study}, the major increase in the number of articles promoting this narrative occurred in the weeks prior to the Russo-Ukrainian War (specifically jumping in volume on February 8, 2022). The most prominent websites that pushed this narrative were unsurprisingly known Russian propaganda outlets including {globalresearch.ca}~(68~articles), {sputniknews.com}~(55), and {rt.com}~(46). Beyond these known pro-Russian websites, we find US-based websites like {veteranstoday.com}~(64~articles), {sott.net}~(62), and {thegatewaypundit.com}~(21) repeating this narrative.  


\vspace{3pt}
\noindent \textbf{Killer Covid-19 Vaccines} (Keywords: Vaccine, Safe, Adverse, MRNA, Effect): 
One prominent misinformation narrative about COVID-19 that we identify is that COVID-19 vaccines are ``killer vaccines'' and a major cause of death around the world. 
For example on lewrockwell.com, an author wrote:\footnote{\url{https://web.archive.org/web/20220120061509/https://www.lewrockwell.com/2022/01/vasko-kohlmayer/dangerous-and-deadly-over-1000-scientific-studies-referencing-injuries-and-deaths-from-covid-vaccines/}}
\vspace{1pt}
\begin{displayquote}
\small
\textit{Whatever they may be, these vaccines are most definitely not safe. We can very clearly see this from the explosion of reports of death to the Vaccine Adverse Event Reporting System (VAERS), which coincided with the introduction of the Covid injections in late 2020.}
\end{displayquote}
\vspace{1pt}

\noindent
As seen in Figure~\ref{fig:case-study}, stories about ``killer vaccines'' have remained prominent throughout 2022, increasing in popularity several times throughout the year. The sites that most prominently echoed this narrative were {theepochtimes.com} (53~articles), {pandemic.news}~(36), and {vaccines.news}~(31). This is consistent with prior studies~\cite{Perrone2022, Daigle2021}.

\vspace{3pt}
\noindent \textbf{2020 Election Denialism} (Keywords: Fraud, Election, 2020, Irregular, Voter): 
The narrative that the presidential election was stolen and that current President Biden is illegitimate~\cite{Baker2020} spread throughout social media and was a key aspect of the January 6, 2021 attack on the US Capitol~\cite{hanley2022no}. We see in our dataset that this false narrative maintained a substantial presence amongst {unreliable} news sites (Figure~\ref{fig:case-study}). For example, the fringe website {thetrumpet.com} wrote on January 6, 2022:\footnote{\url{https://web.archive.org/web/20220107091002/https://www.thetrumpet.com/stephen-flurry/25070-the-insurrection-hoax-is-a-cover-up-for-the-stolen-election}}
\vspace{1pt}
\begin{displayquote}
\small
\textit{The insurrection hoax is a cover-up for the stolen election}
\end{displayquote}
\vspace{1pt}

\noindent The websites that most prominently repeated this narrative were {welovetrump.com}~(143~articles), {thegatewaypundit.com} (55), and {votefraud.news}~(14). 

\subsection{Communities of Unreliable News Websites\label{sec:unreliable}}

To begin to capture the semantic communities that exist within the {unreliable} news ecosystem, we utilize each website's distinct distribution of articles among our discovered set of 52,036~stories/narratives. To compare each website's reporting choices and semantic content, we represent each website's narratives as a multinomial distribution. For example, if we had three narratives (rather than 52.0K) and a website that wrote five articles about Narrative 1, four articles about Narrative 2, and one article about Narrative 3, the website's distribution would be [0.5, 0.4, 0.1]. We do this for all 52,036 narratives and 1,334 websites, thus representing each website as a 52,036-dimensional vector of probabilities. We then use Jensen-Shannon Divergence~\cite{nielsen2020generalization} (detailed in Appendix~\ref{sec:js-divergence}) to compare websites' probability vectors. For example, the JS-Divergence of {rt.com} and {sputniknews.com.com}, two Russian state-sponsored websites~\cite{hanley2022happenstance,center2020pillars} is 0.412 while the JS-Divergence of {rt.com} and {nypost.com}, a US-based website, is 0.605.

After calculating each website's narrative similarities using JS-Divergence with every other website in our dataset, we build an undirected graph with edge weights based on these values (\textit{i.e.,} an edge between a website $P$ and $Q$ is given a weight of $1-JS(P||Q)$) where $JS(P||Q)$ is the JS-Digerence between websites  $P$ and $Q$. We determine communities of websites using the Louvain clustering algorithm~\cite{que2015scalable}. Louvain clustering identified 3~communities, and from these communities,  we qualitatively identified the corresponding three semantic communities: \textit{US-focused}, \textit{International}, and \textit{Conspiratorial}. We label these clusters based on the top topics found within each cluster, with the US-focused cluster writing about Abortion and the Biden Administration, the International cluster about the Russo-Ukrainian War, and the Conspiratorial cluster heavily writing about COV1D-19 vaccines.

\vspace{2pt}
\noindent
{\textbf{US-focused Community.} 696~websites fall into our US-focused community including sites like {dailywire.com}, {breitbart.com}, and {welovetrump.com}. The most common narrative in the community concerned the US Supreme Court Dobbs v.\ Jackson decision to overturn the 1973 Roe v.\ Wade decision that provided the federal right to abortion (Keywords: Roe, Abortion, Wade, Overturn, 1973). To further examine the role of this website community, particularly in regard to its most prominent narrative, we collect a larger set of narratives that more broadly relate to the topic of abortion by aggregating all 121~narrative clusters whose centers have a 0.50~similarity to the Abortion/Roe cluster. 

We consider a website to \textit{originate} a narrative if they published an article about the narrative on the first day that the narrative appeared in our dataset (more than one website can originate a narrative). Altogether, we find that 66.1\% of Roe/Abortion narratives originated from this community, with the website {rawstory.com} originating the most Roe/Abortion narratives (24). In addition to originating most of the narratives about abortion, these websites contributed 77.0\% of the articles on the 121~narratives about abortion; {theepochtimes.com} (1,330 articles) and breitbart.com (1,264~articles) had the most. Largely expected, many International websites such as dailymail.co.uk (1,104 articles) and Conspiratorial websites like evil.news (65 articles) also picked up on these US-centered political narratives, evidencing the spread of stories from this community. 

\vspace{2pt}
\noindent {{\textbf{International Community.}} 405~websites fall into our International community including {rt.com} and {dailymail.co.uk}. The top story was one of our top overall narratives: the Russian invasion of Ukraine (Keywords: Ukraine, Kyiv, Troop, Russian, Donbas). We gather a larger set of 432~narrative clusters that discuss the Russo-Ukrainian War using the same methodology outlined in the prior section. 

We find that 42.4\% of Russo-Ukrainian War narratives started from the International community of websites with 4.6\% of these narratives specifically starting on nine pro-Russian propaganda websites~\cite{center2020pillars}. Globalresearch.ca (19~narratives, tt.com (9~Ukraine~narratives) and tass.com (10~Ukraine~narratives) originate the most narratives among these Russian websites. We again find that this cluster of websites is responsible for a large portion (56.12\%) of articles about the war. Again, largely expected, other websites such as nypost.com (2,934~articles) or treason.news (32~articles) write extensively about the conflict as well. 

\vspace{2pt}
\noindent {\textbf{Conspiratorial Community.}}
233~websites belong to our Conspiratorial community, including popular sites known for spreading conspiracy theories about QAnon and COVID-19~\cite{hanley2022no,hanley2022golden} like
{unz.com}, {qresear.ch}, and {radiopatriot.net}. Unsurprisingly, the top narrative within this community concerns COVID-19 (Keywords: Children, Risk, Adult, Covid, Immunity). We gather a more extensive set of narrative clusters that discuss the COVID-19 and/or COVID-19 vaccines using the same methodology outlined before; altogether gathering 146 narratives. Most prominently, the website childrenshealthdefense.org, the nonprofit run by Robert F. Kennedy Jr., wrote about nearly every COVID-19 story in our dataset (1,774~articles).

We find that our set of Conspiratorial websites originate 38.3\% of narratives about COVID-19, the most prominent of these being {nvic.org}  (15~COVID~narratives) and {covidreference.com} (11~COVID~narratives).  We note that COVID-19 narratives originated not only from these websites but from our International (28.0\%) and US-focused cluster (33.3\%) as well. However, despite this, we find that this cluster is only responsible for 17.8\% of the articles about COVID-19. This shows that COVID-19 narratives came from multiple sources and spread throughout the misinformation news ecosystem.

\section{Originating and Amplifying Narratives}
\begin{table*}
\begin{minipage}{.47\textwidth}
\centering
\fontsize{7pt}{5pt}
\setlength{\tabcolsep}{4pt}
\selectfont
\begin{tabular}{llrlrl}
\toprule
 & CrUX &  Wtd. Ext.   &   Cohen's&  To Peak   & Cohen's \\
Domain& Rank &  Art.$\Delta$   &   D&   $\Delta$  (Days) &  D\\ \midrule
dailymail.co.uk&  $<$ 1K &$0.301$ & $0.679$&-$20.23$ & -$0.376$ \\
express.co.uk & $<$ 1K & $0.161$& -$0.026^*$& -$7.86$&$0.040^*$\\

breitbart.com &1K--5K & $0.388$& $1.075 $&-$54.89$ &-$0.877$ \\
nypost.com&  1K--5K & $0.362$&  $0.819$ &-$45.47$ &-$0.736$  \\
zerohedge.com &1K--5K &$0.180$ &  ${0.628}$ &-$34.18$  &-$0.614$ \\

thegatewaypundit.com &5K--10K &$0.300$&  $1.045$ &-$62.32$  & -$0.980$\\
newsmax.com&5K--10K &$0.306$&  $0.832$ &-$32.58 $ &-$0.690$ \\
dailystar.co.uk &5K--10K &$0.193$&  $0.213^*$ &-$41.12$  &-$0.540$ \\

redstate.com & 10K--50K & $0.502$&$1.413$  &-$69.0$ &-$1.129$ \\
twitchy.com& 10K--50K &$0.454 $&$1.390$& -$71.89$ &-$1.318$\\
dailywire.com& 10K--50K &$0.453$ & $1.345$ &-$73.45$ &-$1.316$ \\

theconservativetreehouse.com& 50K--100K & $0.628$ &$1.969$& -$79.13 $&-$1.612$\\
halturnerradioshow.com& 50K--100K & {$0.475$} & $1.578 $&-$51.22$  & -$0.987$\\
justthenews.com & 50K--100K &$0.811$ &$1.206$& -$71.62$&-$1.218$\\
--\phantom{None}&--&--&--&--&--\\
\bottomrule

\end{tabular}
\end{minipage}
\hspace{25pt}
\begin{minipage}{.49\textwidth}
\centering
\setlength{\tabcolsep}{4pt}
\fontsize{7pt}{5pt}
\selectfont
\begin{tabular}{llrlrl}
\toprule
 & CrUX &  Wtd. Ext.   &   Cohen's&  To Peak   & Cohen's \\
Domain& Rank &  Art.$\Delta$   &   D& $\Delta$  (Days) &  D\\ \midrule
therightscoop.com & 100K--500K &$0.684 $& $1.758$& -$82.53$& -${1.791}$\\
weaselzippers.us & 100K--500K &$0.674$ &$1.519$ & -$81.88 $&${1.786}^*$\\
toddstarnes.com & 100K--500K &$0.576$ &$1.475$ &-$70.96$ &-$1.434$\\

nationalfile.com&  500K--1M &$0.406$&  $1.306$&-$70.21$ & -$1.341$\\
gellerreport.com& 500K--1M&$ 0.359$ &$1.259$ & -$141.96$ &-$2.279^*$\\
ussanews.com&  500K--1M& $0.300$ &  $1.238$& -$121.92$ &-$2.567$\\

infostormer.com& 1M--5M& {$0.763$}& $\mathbf{2.514}$ & {-$\mathbf{154.65}$} &-$\mathbf{4.212}$\\
projectveritas.com&  1M--5M&$0.617$& $2.112$ &{-$62.98$}  &\textbf{-$1.199$}\\
pacificpundit.com &  1M--5M &$0.711$&  ${1.711}$ &-$11.10$  &-$2.139 $\\

anonhq.com&  5M--10M& $\mathbf{1.152}$&  $1.614$& -$93.18 $&-$2.703$\\
redstatenation.com&  5M--10M& $0.502$ &  $1.219$&-$71.13$&-$1.218$\\
thejeffreylord.com &  5M--10M&$0.554$ & $1.185$ &-$61.55$ &-$1.133$\\

thefreedomtimes.com&  10M--50M& $0.530$ & $1.610$ &-$58.44$&-$1.174$ \\
presscorp.org&  10M--50M&$0.554$& $1.460$& -$85.46$ &-$1.462$ \\
trueviralnews.com&  10M--50M&{$0.218$}&  $0.598$&-$52.85$ &-$0.746$\\
\bottomrule

\end{tabular}
\end{minipage}
\caption{ We present the weighted average change (and effect-sizes) in the number of external articles that are published by a random subset of 100 external domains in the week after the website publishes the narrative (\textit{i.e.,} articles not written by the origin domain) and the average change in time (and effect-sizes) for a story to peak in popularity when a website originates a narrative.  We utilize the Mann-Whitney U-test for significant differences in the means. After applying the Bonferroni correction, we conclude that a value is significant if the p-value is $ <0.0017$ (\textit{i.e.}, 0.05/29). We star values that are \emph{not significant}. }
\label{tab:influence-test}
\vspace{-5pt}
\end{table*}

\begin{table*}
\begin{minipage}{.48\textwidth}
\centering
\setlength{\tabcolsep}{3pt}
\fontsize{7pt}{5pt}
\selectfont
\begin{tabular}{llrlrl}
\toprule
 & CrUX &  Wtd. Ext.   &   Cohen's&  To Peak   & Cohen's \\
Domain& Rank &  Art.$\Delta$   &   D&    $\Delta$  (Days) &  D\\ \midrule
dailymail.co.uk & $<$ 1K &$0.626$& $1.546$  &-$14.59$ &-$0.102 $\\
express.co.uk &  $<$ 1K &$0.513$ &  $0.745$&-$16.52$&-$0.187 $\\

breitbart.com&1K--5K &$0.821$ &$1.726$&-$21.73$  & -$0.176 $\\
nypost.com&1K--5K & $0.739$& $1.617$  & -$16.55$& -$0.076$ \\
zerohedge.com&  1K--5K &$ 0.531$& $1.188$ &-$28.57$ & -$0.291$\\

thegatewaypundit.com&5K--10K &$0.754$& $1.519$&-$23.13$ &-$0.192$\\
newsmax.com &5K--10K &$0.649$&  $1.331$ &-$25.75$ &-$0.261 $ \\
rawstory.com&5K--10K &$0.540$&$1.080$&-$16.03$  &-$0.192 $\\

redstate.com& 10K--50K &$0.657$&$1.955$&{-$21.22$} &-$0.171$ \\
babylonbee.com& 10K--50K & $1.726$ & $1.750$&-$31.20$ &-$0.380^*$\\
twitchy.com& 10K--50K & $1.111 $&$1.698$&-$18.52$  &-$0.095$\\

rumormillnews.com & 50K--100K &{$1.427$}& {$2.035$} &-$17.12$ &$0.164^*$\\
beforeitsnews.com & 50K--100K & $0.894$&  $1.989 $&-$32.46$&-$0.312$\\
brighteon.com & 50K--100K &$0.520 $& $1.700$ &-$40.79$ &-$0.479$\\
--\phantom{None}&--\phantom{50M}\\ \bottomrule
\end{tabular}
\end{minipage}
\hspace{12.5pt}
\begin{minipage}{.49\textwidth}
\centering
\setlength{\tabcolsep}{4pt}
\fontsize{7pt}{5pt}
\selectfont
\begin{tabular}{llrlrl}
\toprule

 & CrUX &  Wtd. Ext.   &   Cohen's&  To Peak   & Cohen's \\
Domain& Rank &  Art.$\Delta$   &   D&   $\Delta$  (Days) &  D\\ \midrule
populistpress.com & 100K--500K &{$1.425$} & {$1.861$} &-$\mathbf{43.32}$&-$0.389^*$ \\
henrymakow.com & 100K--500K &$1.148$&$1.831$ &-$17.18$& -$0.190^*$\\
sgtreport.com& 100K--500K &$0.593$ &$1.739$ & -$27.24$ &-$0.320$\\

ussanews.com&  500K--1M & $0.972 $&$2.084 $&-$9.08$ &-$0.113$ \\
politicalflare.com & 500K--1M& $\mathbf{2.011}$ & $1.917$ &-$1.86$&-$0.022^*$\\
yournews.com & 500K--1M& $1.064$ & $1.762$ &-$39.37$&-$\mathbf{0.434}$\\

americafirstreport.com & 1M--5M&$0.611$& $\mathbf{2.406}$ &-$16.84$&-$0.157$  \\
survivethenews.com&  1M--5M&{$0.897$}&  $1.890$ & -$26.40$& -$0.260$\\
barenakedislam.com &  1M--5M & $1.518$&  $1.670$ &-$32.95$ &-$0.207$  \\

patriotjournal.org &  5M--10M& ${1.129}$& $1.864$&-$15.68$ &-$0.064^*$\\
legitgov.org &  5M--10M&$1.156$&  $1.530$ &-{$26.13$}&-{$0.263^*$}\\
gopdailybrief.com &  5M--10M&$0.813$& $1.444$ &-$31.51 $&-$0.377$\\

roguereview.net&  10M--50M&$0.976$& $1.378$ & -$20.78$ &-$0.222^*$\\
trueviralnews.com&  10M--50M& $0.815$& $1.310$ &-$27.59$&-$0.315$\\
thefreedomtimes.com&  10M--50M& $0.641$& $1.187$& -$14.31$ &-$0.006^*$ \\

\bottomrule
\end{tabular}
\end{minipage}
\caption{e present the weighted average change (and effect-sizes) in the external articles that are published by a random subset of 100 external domains for a given domain's amplified narratives (\textit{i.e.,} articles not written by the origin domain) and the average change in time (and effect-sizes) for a story to peak in popularity when a website amplifies a narrative. We utilize the Mann-Whitney U-test for significant differences in the means. After applying the Bonferroni correction, we conclude that a value is significant if the p-value is $ <0.0017$ (\textit{i.e.}, 0.05/29). We star values that are \emph{not significant}.} 
\label{tab:echo-influence}
\vspace{-15pt}
\end{table*}

As seen throughout the last section, several websites play dominant roles in perpetuating and promoting certain types of stories. In this section, we identify and quantify which websites have pivotal roles in originating and amplifying narratives throughout the ecosystem of {unreliable} websites. As before, we consider a website to \textit{originate} a narrative if it published an article about the narrative on the first day that the narrative appeared in our dataset (more than one website can originate a narrative). We consider a website to have \textit{amplified} (\textit{i.e.}, increased the popularity of) a narrative if it (1) posted an article about the narrative before the narrative peaked in popularity, (2) did not originate the narrative, {and} (3) if the posted article appeared in the first 15\% of the total volume of that given narrative. We utilize the 15\% cutoff as it ensures that the vast majority of a narrative's articles have not been published yet (\textit{i.e.}, the story has not dramatically increased in popularity already), allowing us to observe how amplification affects the narrative's popularity. This threshold is consistent with prior work~\cite{leskovec2009meme}. 

With this approach, we investigate how the popularity of a website influences its effectiveness in originating and amplifying narratives using website rank data provided by the Google Chrome User Report (CrUX) from October 2022, which Ruth et~al.\ showed to be the most reliable website popularity metric~\cite{ruth2022toppling}.

\subsection{Originating Narratives\label{sec:origin}}
To measure the efficacy of websites in originating narratives, we perform a correlational comparison of the number of external non-origin articles that are written about a given narrative in the week after a given website originated a narrative vs. the number of non-origin external articles that are written in the week after origination if the website did not originate the narrative but still eventually wrote about that narrative. We note that for this analysis, we weight the number of articles by the log inverse of its CrUX popularity ranking~\cite{ruth2022world, jarvelin2002cumulated} to ensure that we do not consider an article from a highly popular website such as breitbart.com the same as from a relatively obscure website such as welovetrump.com. 

To ensure each website has a marked effect on the full {unreliable} news ecosystem and to improve the robustness of our approach, we utilize a bootstrapping procedure~\cite{little2019causal} ($B=250$) to measure the influence of each website by taking a random subset of 100~websites in each bootstrap and then measuring the weighted increase in the number of articles across this set of a random set of 100~websites. We provide the average effect size (Cohen's~D)/statistical measure of the increase in articles and the p-value to check for the significance (using Mann-Whitey U-tests) for the change in the number of external articles in Table~\ref{tab:influence-test}. For this section, we limit our analysis to websites that consistently originate articles by only considering websites with at least 25~instances of \textit{originating} an article. 

We observe only a small correlation (Pearson correlation $\rho = 0.229$) between a website's popularity and its ability to originate and perpetuate narratives amongst other {unreliable} news websites. For example, considering {express.co.uk} and {dailymail.co.uk}, two tabloids known to engage in sensationalism and biased reporting with the highest CrUX popularities, while {dailymail.co.uk} is fairly effective at originating narratives (Cohen's D = 0.679),  {express.co.uk} is one of the worst at originating new narratives (Cohen's D = -0.026). Further, a seemingly unpopular website, {infostormer.com}, is one of the best at propagating narratives it originates to other sites.  This illustrates many different types of websites can originate and propagate narratives in the misinformation news ecosystem. {Infostormer.com}, with a header labeled the \textit{``Jewish Problem''}, writes heavily sensationalist and antisemitic perspectives on the news that is taken up by other websites.  For example, after writing an article on how the CNBC news host Jim Cramer was promoting Meta stock,\footnote{\url{http://web.archive.org/web/20221028232854/https://infostormer.com/jew-jim-cramer-cries-and-apologizes-for-hyping-metas-stock/}} this news story was later covered by more popular websites like {activistpost.com}\footnote{\url{https://web.archive.org/web/20221028014725/https://www.activistpost.com/2022/10/the-big-tech-companies-are-telling-us-exactly-where-the-economy-is-headed-in-2023.html}} and {hannity.com}.\footnote{\url{http://web.archive.org/web/20220901000000*/https://hannity.com/media-room/sad-money-jim-cramer-in-tears-after-meta-stock-nosedives-i-made-a-mistake/}} 

In addition to quantifying each website's efficacy in originating narratives, we determine how quickly after a website originates a narrative that the story peaks in popularity. Here, a negative Cohen's D indicates that the ``time for a narrative to peak'' occurs faster. This metric, combined with the previous metric, describes how effective a website is at reorienting online conversations to its own narratives. We see only a slight correlation ($\rho=0.164$) with a website's CrUX-defined popularity. Rather, we again see that several small websites are highly effective in originating narratives that peak quickly (\textit{i.e.}, writing about narratives that become of immediate interest). For example, when the small website {infostormer.com} originates narratives, those narratives peak in popularity 155~days earlier than when {infostormer.com} does not originate narratives (Table~\ref{tab:influence-test}). We similarly observe that the right-wing and conspiratorial websites ussanews.com and gellerreport.com, which often wrote about QAnon~\cite{hanley2022no} are also highly effective at quickly landing their narratives on other websites, with some of the lowest ``time to peak'' in our dataset. 

\subsection{Amplifying Narratives\label{sec:magnify}} 

To understand how effective websites are at amplifying narratives, we correlationally compare the number of external non-origin articles that are written about a given narrative when a given website amplifies the narrative versus when the website does not amplify the narrative but still eventually wrote about that narrative. We utilize the same weighting and bootstrapping procedure as in the previous section, again limiting our analysis to websites that amplify at least 25 narratives across our period of study. 
We again observe only a slight correlation between a  website's popularity and its ability to amplify narratives (Pearson correlation $\rho$ = 0.302). As before, we see in Table~\ref{tab:echo-influence} that websites across different CrUX popularities excel at amplifying narratives. For example, One of the most effective websites is {barenakedislam.com}, an anti-Islam website with the slogan \textit{``It isn't Islamaphobia when they really ARE trying to kill you.''} For example, after echoing a narrative about how Muslim men were targeting Ukrainian refugees,\footnote{\url{https://web.archive.org/web/20220422091130/https://barenakedislam.com/2022/03/21/what-a-surprise-not-ukrainian-female-refugees-say-they-dont-feel-safe-in-multicultural-sweden/}} this news story traveled to an additional 12~other {{unreliable}} news websites including more popular websites like americanthinker.com\footnote{\url{http://web.archive.org/web/20220619083901/https://www.americanthinker.com/articles/2022/06/the_only_rape_where_the_left_says_victimblaming_is_okay.html}} and breitbart.com.\footnote{\url{https://web.archive.org/web/20220527122734/https://www.breitbart.com/europe/2022/05/27/sweden-asylum-home-tells-ukrainian-women-dress-modestly-to-not-provoke-migrant-men/}}  This illustrates how different types of websites can amplify and propagate narratives in the misinformation news ecosystem.

Finally, we determine the effect of narrative amplification by each website on how quickly the narrative peaks. There is again a small correlation between website popularity and amplification (Pearson correlation $\rho$ = 0.152). As seen in Table~\ref{tab:echo-influence}, the most effective website at quickly amplifying narratives to their peak popularity is {populistpress.com}, a drudge-style news website that hosts hyperlinks to different news articles. Despite not hosting many articles itself, we see that when it does mention a narrative, this news story is more likely to more quickly peak in popularity.

\begin{figure}
  \centering
  \includegraphics[width=.95\columnwidth]{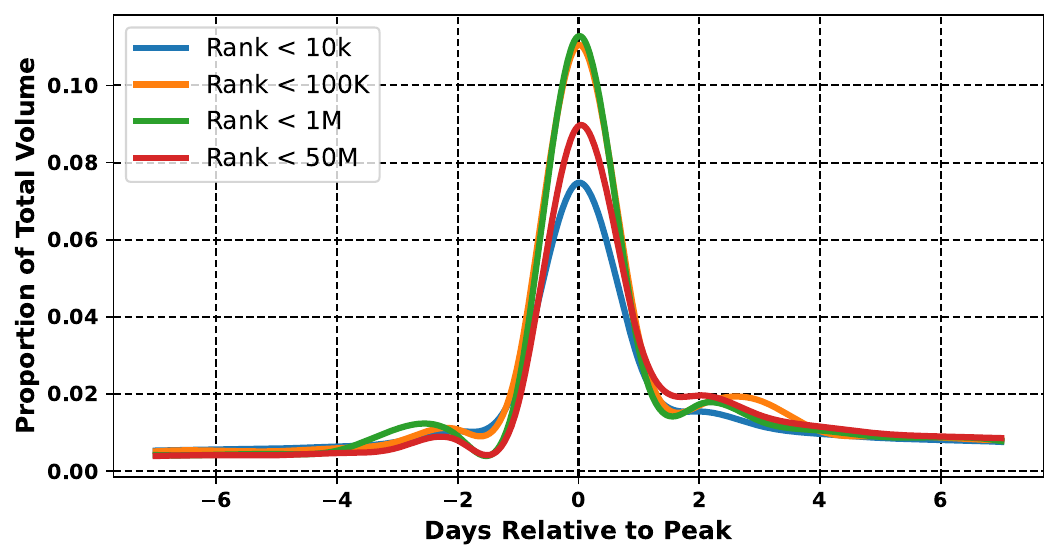}
\caption{Time lag for differently ranked websites. The most popular websites write more of their articles prior to the peak of a narrative's popularity. In contrast, less popular websites tend to respond to narratives and write most of their articles after the peak.}
\label{fig:time-lag}
\vspace{-10pt}
\end{figure}

\vspace{2pt}
\noindent
\textbf{Trend Setting.} 
Despite not seeing clear discernible patterns in how website popularity correlates with a website's ability to originate and amplify narratives, we do observe differences in \emph{when} these websites write about given narratives. As seen in Figure~\ref{fig:time-lag}, across all narratives, more popular websites tend to write fewer articles (as a proportion) on the day that a given narrative peaks. A slightly higher percentage (37.37\%) of articles from websites with a CrUX rank $<$10K come before the peak versus the 31.39\% of articles with a rank above 1M\@. This indicates, as also found by Leskovec et~al.~\cite{leskovec2009meme}, popular websites have \emph{some} ability to set the agenda for the topic smaller websites write.

\subsection{The Role of Fringe Forums/Social Media}

We now analyze the relationship between our set of {unreliable} news websites and the fringe social media sites 8kun and 4chan. As with our set of {unreliable} news websites, we scrape 8kun and 4chan \texttt{/pol} posts published between January 1 and November 1, 2022 (Section~\ref{sec:preprocess}). 8kun data is readily available on their website {8kun.top}; 4chan \texttt{/pol} posts are archived through the website {archive.4plebs.org}. 
Altogether, we gather 632,091~posts from 8kun.top and 4.69~million posts from 4chan (Table~\ref{tab:num-social-media-posts}). 

\begin{table}
\centering
\footnotesize
\begin{tabular}{lrr}
\toprule
 Social Media &  & Posts With Corresponding \\
 Platform & Posts &  News Article Narrative  \\\midrule
8kun.top & 632,091 & 34,959  (5.53\%) \\
4chan.org & 4,690,669 & 450,027 (9.59\%) \\
\bottomrule
\end{tabular}
\caption{Our dataset of social media posts and their relationship to the narratives published by unreliable news websites.} 
\label{tab:num-social-media-posts} 
\vspace{-7pt}
\end{table}

\begin{table}
\centering
\scriptsize
\selectfont
\begin{tabular}{lr}
\toprule
Top Narratives on 8kun &  Comments\\ \midrule
ukraine, nato, putin, russia, war  &625   \\
hillary, collusion, clinton, mueller, lie&454 \\
antisemit, jew, israel, zionist, israel&373\\
trucker, trudeau, canadian, ottawa, convoy &365 \\
ukraine, kyiv, troop, russia, donbas &286 \\
\bottomrule
\toprule
Top Narratives on 4chan & Comments\\   \midrule
ukraine, kyiv, troop, russia, donbas& 4,361  \\
volodymyr, zelenskyy, ukrainie, kyiv, president&3,003\\
ukraine, conflict, war, escalation, tension&2,792 \\
race, white, theory, black, crt&2,134 \\
jew, white, supremacy, goyim, zionist&2,079\\
\bottomrule
\end{tabular}
\caption{ The top topics from our set of unreliable news websites present on 8kun and 4chan /pol.} 
\label{tab:social-media-topics}
\vspace{-10pt}
\end{table}


To find the correspondence of 8kun and 4chan comments between news narratives, we preprocess, embed, and assign each 8kun post to its most similar narrative cluster. As before, we utilize a threshold of 0.60 for matching a comment to its corresponding news article narrative. Altogether, we find that 34,959 (5.53\%) of comments on 8kun.top and 450,027 (9.59\%) of comments from 4chan.org  correspond to a narrative on our set of {unreliable} news websites (Table~\ref{tab:num-social-media-posts}).

\subsubsection{8kun}
Examining the top narratives posted on 8kun that correspond with a news narrative (Table~\ref{tab:social-media-topics}), we see that the most commonly shared narratives on 8kun.top concern the Russo-Ukrainian war, the investigation of Donald Trump by Special Counsel Robert Mueller~\cite{Zapotosky2020}, antisemitic beliefs, and the 2022 Trucker Convoy in Ottawa Canada~\cite{Murphy2022}. This largely corresponds with 8kun being known as the home of hard-right, conspiratorial, and antisemitic posts~\cite{Glaser2019}. As in Section~\ref{sec:unreliable}, we determine the distribution of narratives from our set of unreliable websites that are present on 8kun to understand the similarity between 8kun and the collective narratives on our set of {unreliable} news websites. Altogether, 8kun has a JS-Divergence of 0.240 with the collective narrative distribution of {unreliable} news websites (Table~\ref{tab:kun-4lebs-echo}). Performing this on an individual site level, we observe several of the websites with the most similar narrative distributions prominently discuss conspiratorial ideas (\textit{e.g.}, lucianne.com and radiotpatriot.net)~\cite{hanley2022no}.
\begin{table}
\centering
\scriptsize
\selectfont
\begin{tabular}{lrrrrr}
\toprule
& Narratives &Wtd. Ext.  & Cohen's&To Peak & Cohen's \\ 
Platform &  Originated &   Art.$\Delta$ &D  &  $\Delta$  (Days) &D   \\ \midrule
8kun &$392$ &$ \mathbf{0.400}$ & $\mathbf{1.759}$ & -$\mathbf{73.92}$ & -$\mathbf{1.117}$    \\
4chan & $\mathbf{2,455}$ & $0.394$ &$0.842$ & -$29.45$& -$0.434$  \\ \bottomrule \toprule
& Narratives &Wtd. Ext. &Cohen's&To Peak &Cohen's\\
Platform & Amplified &  Art.$\Delta$&  D &  $\Delta$  (Days) &  D  \\ \midrule
8kun &  $2,728$ &$\mathbf{0.760}$&{$0.283$}  &-$\mathbf{19.82}$&-$\mathbf{0.353}$\\
4chan & $\mathbf{12,164}$& ${0.327}$ &$\mathbf{0.913}$& ${1.19}$ & ${0.017}^*$  \\\bottomrule \toprule
  & JS-Sim. & \multicolumn{4}{c}{} \\  
Platform & to News & \multicolumn{4}{l}{Most Similar News Sites} \\\midrule
8kun & $0.240 $& \multicolumn{4}{l}{lucianne.com, radiopatriot.net, americanthinker.com} \\
4chan & $0.248$& \multicolumn{4}{l}{unz.com, beforeitsnews.com, thetruthseeker.co.uk} \\
\bottomrule
\end{tabular}
\caption{ The influence of 8kun and 4chan on the ecosystem of unreliable news websites and their similarity (by JS-Divergence) to the {unreliable} news dataset. We star values were \emph{not found to be significant} according to the Mann-Whitney U-test.  }
\label{tab:kun-4lebs-echo}
\vspace{-15pt}
\end{table}
Having examined similarities between the narratives discussed on 8kun and those on particular websites in our dataset, we next determine the influence of 8kun on our {unreliable} news website ecosystem. We utilize the same definitions of \textit{originate} and \textit{amplify} as well as the same methodology as in Sections~\ref{sec:origin} and~\ref{sec:magnify}.  As seen in Table~\ref{tab:kun-4lebs-echo}, 8kun originating or amplifying a particular narrative has a modest effect on the number of articles written about that narrative. In the week after 8kun originates a given narrative, we see an average Cohen's D of 1.759. In contrast, in the week after 8kun users amplify a narrative, we observe a  Cohen's D of 0.283, illustrating that 8kun is somewhat better at originating narratives than amplifying narratives.  Thus, while not as effective as some of the websites in our dataset (Tables~\ref{tab:influence-test} and~\ref{tab:echo-influence}), when 8kun users comment on narratives, this correlates with a slight increase in the narrative's popularity. We see this further mirrored in the effect that 8kun has in expediting narratives to peak earlier. On average, if 8kun originates a narrative, it peaks in popularity 73.9~days earlier than if 8kun did not originate the narrative. Similarly, if 8kun amplifies a narrative it peaks in popularity 19.8~days earlier on average. 
\subsubsection{4chan /pol}
Looking at the top corresponding shared narratives on 4chan /pol, we see several that target Judaism and the Jewish people (Table~\ref{tab:social-media-topics}). As with 8kun, 4chan has a reputation for antisemitism and racist language. Besides the narratives that center on the Russo-Ukrainian war, we see this racism and antisemitism reflected in the top shared narratives on the website~\cite{zelenkauskaite2021shades}. Determining the distribution of narratives from our set of {unreliable} news websites that are present on 4chan, altogether, 4chan has a JS-Divergence of 0.248 with the collective narrative distribution of {unreliable} news websites (Table~\ref{tab:kun-4lebs-echo}). Examining the most similar websites to 4chan, we observe several with known conspiratorial reputations.  As documented by Medias-Bias/FactCheck, the most similar website to 4chan, {unz.com} is a conspiratorial and hate-oriented website that often cites white nationalist groups in its articles~\cite{unz2022}. Similarly, {beforeitsnews.com}~\cite{before022} and {thetruthseeker.co.uk}~\cite{truthseek2022} are known to ``promote conspiracy theories and pseudoscience.'' 

Finally, we determine the role 4chan has in promoting and amplifying narratives within our ecosystem of {unreliable} news websites. We observe a similar effect to 8kun, in terms of the weighted increase of articles when 4chan originates a narrative compared to when it does not (Cohen's D of 0.438). However, unlike 8kun, we do observe that 4chan is better at amplifying narratives, with an effect size of Cohen's D of 0.913 (Table~\ref{tab:kun-4lebs-echo}). However, in contrast to 8kun, 4chan is relatively less effective at getting the narrative to peak earlier. If 4chan users originate a narrative, it peaks in popularity 29.5 days earlier compared to 73.9 days earlier when 8kun originates a narrative. When 4chan amplifies a narrative, it has little effect on when that narrative peaks.

\section{Detecting Narratives and Fact-Checking\label{sec:fact-checking}}
In the last two sections, we analyzed the narratives and behavior of {unreliable} news websites during 2022. In this section, we present two case studies that highlight how our programmatic approach can also identify new narratives and assist in focusing fact-checking efforts.

\subsection{Identifying New Trending Narratives}
By examining the week-over-week percentage increases in story volumes, we programmatically determine which narratives are receiving new or renewed focus on {unreliable} news websites, which is imperative for ameliorating the spread of specious information~\cite{rajdev2015fake, wu2019misinformation, saeed2022trollmagnifier}. 
The narratives that increased most in volume during the last week of our experiment (October 26 to November 1, 2022) were:

\vspace{2pt}
\noindent
\textbf{The Attack of Paul Pelosi, Keywords: Pelosi, Depap, hammer, Nancy, Paul.} On October 28, 2022, the husband of Congresswoman Nancy Pelosi was attacked in his home~\cite{TimesPelosi2022}. Largely due to the proximity of the time of the attack to the 2022 US midterm elections, the attack became a source of conspiracy theories and wild speculation. For example, one user wrote on thegatewaypundit.com:\footnote{\url{https://web.archive.org/web/20221028130434/https://www.thegatewaypundit.com/2022/10/breaking-pelosis-home-broken-early-morning-san-francisco-paul-pelosi-violently-beaten-taken-hospital/}}

\vspace{1pt}
\begin{displayquote}
\small
\textit{\textit{Whenever bad things happen to Paulie P, his wife always manages to have an alibi.}}
\end{displayquote}
\vspace{1pt}

\noindent 364~articles (compared to zero the week before) were written about the event within our dataset across 175~websites. The {thegatewaypundit.com} had 44~articles, {dailymail.co.uk.com} had 40, and {nypost.com} had 37.

\vspace{3pt}
\noindent
\textbf{The Seoul Halloween Stampede, Keywords: Halloween, Seoul, Itaewon, Festivity, Stampede.} On October 29, 2022, a crowd rush in the Seoul neighborhood of Itaewon resulted in the death of 158~people. Across our dataset, we  see 132~articles across 46~websites written about this event, with 18~articles from {dailymail.co.uk}, 11 from {republicworld.com}, and 8 from {mirror.co.uk}. 

\vspace{3pt}
\noindent
\textbf{Elon Musk's First Visit to Twitter Headquarters, Keywords: Sink, Headquarters, Musk, Carry, Twitter.} After officially purchasing the social media company Twitter, on his first visit to the company on October 26, Elon Musk carried a sink into the headquarters with him. This prop humor by Musk was supposed to be a play on ``let that sink in'' but with a real sink. We see 176~articles from 84~domains about the story, with 10~articles from the {dailymail.co.uk}, 9 from {westernjournal.com}, and 8 from {conservativeangle.com}.

\subsection{Fact-Checking}
One approach to combating the spread of new misinformation stories that many organizations have adopted is fact-checking. Fact-checking a story requires hours to deeply understand its context and nuance~\cite{factcheck-wapo}. Unfortunately, this means that propaganda and misinformation often spread widely in a rapidly evolving media landscape before journalists can respond. Our approach can serve as a way to programmatically identify new misinformation narratives {as they appear} and begin to gain traction, ideally reducing the amount of time from when a story is published to when a fact-checker can respond. 

To show how our system might be useful to fact-checking organizations, we utilize our approach to analyze the behaviors of particular narratives before being fact-checked by three organizations: Politifact, Reuters, and APNews~\cite{zlatkova2019fact,nyhan2010corrections,rashkin2017truth}.  For the three agencies, we gathered the set of fact-checking articles that each published in 2022. Altogether we scraped~1,524, 3,090, and 140~articles from Politifact~\cite{politifactFactCheck2022}, Reuters~\cite{reutersFactCheck2022}, and APNews~\cite{apnewsFactCheck2022}, respectively. To augment our system to perform fact-checking (\textit{i.e.}, determine whether a fact-check article \emph{refutes} a given narrative), we additionally train a DeBERTa-based~\cite{he2022debertav3} classifier on the FEVER~\cite{thorne2018fever} dataset that takes a claim (\textit{i.e.}, an article) and a query (\textit{i.e.}, a fact-check) and labels the query as either \textit{supporting} the claim, \textit{refuting} the claim, or \textit{not having enough information} to say anything about the claim. Using 10\% of the FEVER dataset as a held-out test set, our DeBERTa-based model achieves an overall 90.7\% accuracy on this test set (90.5\% precision in labeling refutations).

\begin{table}
\centering
\setlength\tabcolsep{5pt}
\scriptsize
\begin{tabular}{llllll}
\toprule
& Narr. & Med.    &Med.  &Med. &    \\ 
&{Fact-}  & Art. Prior   & {Days to }  &{Days from }  &{0-Day }   \\ 
&{Checked}  &Fact-Check   & {Fact-Check} &{Narr. Peak } & {Fact-Checks} \\ \midrule
Politifact & 6,231 &6 &55.0  & 4.0 & 110 \\
Reuters & 9,604 &3& 49.0 &0.0 & 647 \\
AP News & 230&  15 &83.0 & 3.0&8 \\
\bottomrule
\end{tabular}
\caption{ Efficacy of fact-checking websites. All three websites most commonly fact-check---by the number of articles with the same narrative as the fact-check---the articles of resear.ch (433~articles), gatesofvienna.net (414), dailymail.co.uk (406).} 
\label{tab:fact-check}
\vspace{-15pt}
\end{table}

For each fact-checking article, as with articles from unreliable websites, we divide the article into its constituent passages and embed them utilizing our MPNet model. We consider a narrative to have been addressed by a fact-checker if the fact-checker writes about the narrative.  We note that articles frequently ``fact-check'' or ``add context'' to multiple narratives.  To provide fact checkers with the greatest number of ``opportunities'' to fact-check a narrative, we map each fact-check passage to \emph{all} articles above our cosine similarity threshold of 0.60 rather than map the fact-check passage to only the single closest narrative. After mapping these fact-checking passages to our set of articles, we utilize our DeBERTa-based fact-checking classifier to ensure that the corresponding ``fact-check'' \emph{refutes} the information of the corresponding unreliable news article passage.  We provide an example of an identified fact-check below. To ensure that our model is able to properly identify ``fact-checks'', we manually validate 100~random fact-check-article refutation pairs, finding that 94\% of them are indeed refutations. 
\vspace{5pt}

\noindent\fcolorbox{black}{lightgray}{%
\begin{minipage}{.47\textwidth}
\scriptsize
\textbf{Article Passage:} This is a shining example and a small part of why it is so vitally important to find the underlying cause of the fraud that took place both in November 2020 and the lead-up to that election. 
\vspace{1pt}\\
\textbf{Fact-Check:} THE FACTS: To be clear, no widespread corruption was found and no election was stolen from Trump.
\end{minipage}}
\vspace{3pt}

As seen in Table~\ref{tab:fact-check}, on average, narratives can spread one to two months before being fact-checked by these reputable websites. Furthermore, on average both Politifact and AP News write about stories after they have peaked in popularity; AP News, with the fewest fact-checks, writes about narratives right as they peak in popularity. We further see a heavy overlap between the narratives that each website fact-checks. Reuters and Politifact have an overlap of 2,646~stories/narratives; AP News and Politifact, have an overlap of 137~narratives; and Reuters and AP News have an overlap of 141~narratives. Furthermore, the unreliable websites that have articles most commonly fact-checked by the three fact-checking organizations are the same: qresear.ch (433~articles), gatesofvienna.net (414), and dailymail.co.uk (406). This underscores that these fact-checking websites are duplicating effort, often fact-checking the same narratives~\cite{graves2018understanding}. We note, however, that while narratives often spread for long periods before being fact-checks, the number of articles, on average, is often low~(3--15~articles). We thus see that many fact-checkers \emph{are} effective at fact-checking narratives when they peak in popularity, but often understandably do not fact-check narratives that have just begun to spread among different {unreliable} news websites.

We see that many narratives spread for long periods on {unreliable} news websites before they are fact-checked near their narrative peak. However, our system can surface these narratives to fact-checkers long before they peak in popularity, aiding in the fact-checking organizations' typical workflow in identifying potential misinformation. This can enable fact-checkers to identify and address misleading narratives concurrent to when they first rise in popularity. 



\section{Related Work}

Our study builds on considerable prior work on both the spread of misinformation online and language models. There have been several past quantitative studies of the spread of information online.
Leskovec et~al.\ identify the trends in the propagation of ``memes''~\cite{leskovec2009meme}. They find that while the majority of memes originate from mainstream websites, key phrases that start on smaller blogs are often adopted by larger platforms. Gomez-Rodriguez et~al.\ adopt a cascade transmission model and identify how best to estimate the relative influence of different news outlets in spreading stories~\cite{gomez2012inferring}. Similar to our use of DP-Means, prior works have utilized CluStream among other clustering techniques to track information or news over time~\cite{tajalizadeh2019novel,alsayat2016social,fan2021clustering}. For example, Curiskis et~al.~\cite{curiskis2020evaluation} utilize document clustering based on dictionaries to track topics. 

\vspace{2pt}
\noindent
\textbf{Analyzing the Spread of Misinformation.}
Several works have tracked the spread and impact of misinformation. Shu et~al.~\cite{shu2017fake} present the largest overall overview of misinformation detection issues, presenting various paradigms for tracking and labeling misinformation. These include tracking news content features and social content features. For example, Cao et al.~\cite{cao2020exploring} and Meel et al.~\cite{meel2020analysing} explore utilizing image and text-based features to label misinformation. Abdali et al., in contrast, use screenshots of websites to identify the trustworthiness of websites and label misinformation~\cite{abdali2021identifying}. Extensive work has studied individual campaigns that spread unreliable information, on topics like QAnon~\cite{hanley2022no, hanley2022golden}, Syrian White Helmets~\cite{starbird2018ecosystem}, the Russo-Ukrainian War~\cite{hanley2022happenstance,hanley2022special}, and COVID-19~\cite{madraki2021characterizing}. 

Recent work from the security community has focused on identifying and curbing misinformation. Kaiser et~al.~\cite{kaiser2021adapting} studied how borrowing techniques from the security warning landscape might help to inform mis/disinformation warnings. Paudel et~al.~\cite{paudel2023lambretta} recently demonstrated how techniques like Learning To Rank (LTR) can be used to soft-moderate misinformation on Twitter. On the human-level, Sharevski et~al.~identified folk models of misinformation on social media that could inform potential defenses~\cite{sharevski2022folk}.

\vspace{2pt}
\noindent
\textbf{Language Models, Semantic Search, and Topic Analysis.}
Many previous topic analysis methods have been built on Latent Dirichlet Allocation (LDA). Albalawi et~al.\ show that LDA is one of the most effective methodologies for extracting topics from short text data compared to other computationally light alternatives proposed within the last decade (\textit{e.g.}, LSA, LDA, NMF, PCA, RP)~\cite{albalawi2020using}. Meng et~al.~\cite{meng2022topic}, Angelov~\cite{angelov2020top2vec}, and Grootendorst~\cite{grootendorst2022bertopic} have enabled users to perform topic modeling utilizing large language models. Utilizing these techniques and online document clustering~\cite{yin2018model,blei2010nested}, others have performed robust, but smaller scoped semantic analysis (e.g., on Russian disinformation campaigns~\cite{hanley2022happenstance,hanley2022partial}).



\section{Discussion and Conclusion}
In this work, we introduced and validated a new, scalable methodology for tracking news narratives online. Applying the methodology to study the stories published on 1,334~unreliable news websites during 2022, our work shows how a large-scale, quantitative analysis can identify propagation patterns and significant players that may otherwise have been difficult to uncover through qualitative investigations of individual disinformation campaigns. Specifically, we showed that less frequented websites and fringe social media platforms can have marked effects on amplifying the narratives discussed on unreliable news websites. 

Our study also highlights the need to programmatically detect the rise of false narratives in real time. Prior work has shown that misinformation can spread ten times faster than legitimate news~\cite{vosoughi2018spread} and our analysis finds that false narratives can often start on small, seemingly unpopular websites. In many cases, these false narratives spread for months online before being fact-checked. As such, we are exploring how best to publicly and continuously release real-time updates of our narrative analyses on an online dashboard while protecting against misuse (\textit{e.g.}, use in AI training models~\cite{David2023} or targeted misinformation chatbots).

While our study illustrates the potential for programmatically tracking news narratives, it also simultaneously surfaces areas for further research. For example, as found in past works~\cite{hanley2022happenstance,isaac2021ethical}, though our approach can identify precise stories/narratives within our dataset, medical information poses challenges for large language models like MPNet. For example, one of the misclassifications (Narrative~6 
in Table~\ref{tab:precision-test}) concerned COVID-19 rather than Monkeypox. Given the high knowledge level needed in understanding medical misinformation, past works have recommended utilizing models specifically trained for medical misinformation for topic analysis of these stories~\cite {isaac2021ethical}. We hope that the potential of and demonstrated need for programmatic approaches for tracking news narratives and misinformation online motivates further work on these topics.

\ifCLASSOPTIONcaptionsoff
  \newpage
\fi



\bibliographystyle{IEEEtran}
\bibliography{IEEEexample}
%

\appendices
\newpage
\section{Training with Unsupervised Contrastive Loss \label{sec:contrastive}}

To train our MPNet model, we utilize unsupervised contrastive learning to better the quality of our embeddings~\cite{gao2021simcse}. For training, this is such that we embed each example $x_i = (text_i) \in D_{News}$ (where $text_i$ is the text) twice (with dropout both times) using MPNet by inputting $[CLS] text_i [SEP]$ and averaging the contextual word vectors of the resulting output as a hidden vector $\mathbf{h}_i$ and $\tilde{\mathbf{h}}_i$  for $text_i$ as its representations. Then, given a set of hidden vectors $\{\mathbf{h}_i\}_{i=0}^{N_b}$ and $\{\tilde{\mathbf{h}}_{j}\}_{j=0}^{N_b}$ (different dropout), where $N_b$ is the size of the batch, we perform a contrastive learning step on that batch. This is such that for each Batch $\mathcal{B}$, for an \textit{anchor} hidden embedding $\mathbf{h_i}$ within the batch, the set of hidden vectors $\mathbf{h_i} \,, \mathbf{\tilde{h_j}} \in \mathcal{B}$,  vectors where $i = j$ are positive pairs. Other pairs where $i\neq j$ are considered negative pairs. Within each batch $\mathcal{B}$, the contrastive loss is computed across all positive pairs in the batch such that:
\[
    L_{contrastive} = -\frac{1}{N_b} \sum_{\mathbf{h}_i \in \mathcal{B}}\mathit {l}^c(\mathbf{h}_i )
\]
\[
\mathit{l}^c(\mathbf{h}_i) = {log}\frac{ \sum_{j\in\mathcal{B} } \mathbbm{1}_{[i = j]}\mathrm{exp}( \frac{\mathbf{h}_i^\top \tilde{\mathbf{h}_j}}{\tau||\mathbf{h}_i || ||\tilde{\mathbf{h}_j} || })}{\sum_{j\in\mathcal{B}} \, \mathrm{exp}( \frac{\mathbf{h}_i^\top \tilde{\mathbf{h}_j}}{\tau||\mathbf{h}_i || ||\tilde{\mathbf{h}_j} || })}
\]
where, as in prior work~\cite{liang2022jointcl}, we utilize a temperature $\tau=0.07$.
\section{Passage Pairs\label{sec:thresholds}}
\begin{figure}[!h]
\begin{minipage}{.47\textwidth}
\centering\scriptsize
\textbf{0.45 Similarity}
\end{minipage}
\noindent\fcolorbox{black}{lightgray}{%
\begin{minipage}{.47\textwidth}
\tiny
\textbf{PASSAGE 1:} The growing possibility that nuclear weapons might be used, as hostilities in Ukraine continue to escalate, merits your full attention.

\textbf{PASSAGE 2:} Raising the alert level of Russian nuclear forces is a bonechilling development, Guterres declared. The prospect of nuclear conflict, once unthinkable, is now back within the realm of possibility.
\end{minipage}}

\label{figure:all-paragraph_pairs1}
\vspace{-10pt}
\end{figure}

\begin{figure}[!h]
\begin{minipage}{.47\textwidth}
\centering\scriptsize
\textbf{0.50 Similarity}
\end{minipage}
\noindent\fcolorbox{black}{lightgray}{
\begin{minipage}{.465\textwidth}
\tiny
\textbf{PASSAGE 1:} When you actually look at the bill and it says no sexual instruction to kids preK through three, how many parents want their kids to have transgenderism or something injected into classroom instruction? DeSantis said earlier this month.

\textbf{PASSAGE 2:} Parents watchdog group Parents Defending Education PDE has warned that a school district in Minnesota is pushing transgender and pride books and materials on to children as young as three years old.
\end{minipage}}

\label{figure:all-paragraph_pairs2}
\vspace{-10pt}
\end{figure}
\begin{figure}[!h]
\begin{minipage}{.47\textwidth}
\centering\scriptsize
\textbf{0.55 Similarity}
\end{minipage}
\noindent\fcolorbox{black}{lightgray}{%
\begin{minipage}{.47\textwidth}
\tiny
\textbf{PASSAGE 1:} Protests in the Netherlands became violent with police cars being set ablaze as the public grows angry with their enforcement of COVID edicts to restrict their civil liberties:

\textbf{PASSAGE 2:} Thousands of Dutch citizens lined up in the streets defiantly even after government officials banned protest, using the neverending pandemic as an excuse to brutally crackdown on civil liberties.
\end{minipage}}

\label{figure:all-paragraph_pairs3}
\vspace{-10pt}
\end{figure}
\begin{figure}[!h]
\begin{minipage}{.47\textwidth}
\centering\scriptsize
\textbf{0.60 Similarity}
\end{minipage}
\noindent\fcolorbox{black}{lightgray}{%
\begin{minipage}{.47\textwidth}
\tiny
\textbf{PASSAGE 1:} The raid by over 30 plain clothes agents from the Southern District of Florida and the FBI s Washington Field Office extended through the Trump family s entire 3,000squarefoot private quarters, as well as to a separate office and safe, and a locked basement storage room in which 15 cardboard boxes of material from the White House were stored.

\textbf{PASSAGE 2:} Donald Trump lamented Wednesday that the FBI blocked his lawyers from the property during the raid at his Palm Beach, Florida residence and suggested that agents may have 'planted' evidence. 
\end{minipage}}

\caption{Example of passage pairs at different levels of cosine similarities. }
\label{figure:all-paragraph_pairs4}
\vspace{-10pt}
\end{figure}

\section{DP-Means Algorithm\label{sec:dpmeans}}

DP-Means~\cite{kulis2011revisiting} is a non-parametric extension of the K-means algorithm that does not require the specification of the number of clusters \textit{a priori}. Within DP-Means, when a given datapoint is a chosen parameter $\lambda$ away from the closest cluster, a new cluster is formed. Dinari et al.~\cite{dinari2022revisiting} parallelize this algorithm by \textit{delaying cluster creation} until the end of the assignment step. Namely, instead of creating a new cluster each time a new datapoint is discovered, the algorithm instead determines which datapoint is furthest from the current set of clusters and then creates a new cluster with that datapoint. By delaying cluster creation, the DP-means algorithm can be trivially parallelized. Furthermore, by delaying cluster creation, this version of DP-Means avoids over-clustering the data (\textit{i.e.,} only the most disparate datapoints create new clusters)~\cite{dinari2022revisiting}.

\section{Pointwise Mutual Information\label{sec:pmi}}

The PMI of a particular word $word_i$ in a cluster $C_j$ is calculated as:

\vspace{-10pt}
\begin{align*}
\scriptsize
PMI(word_i, C_j) = log_2\frac{P(word_i,C_j)}{P(word_i) P(c_i)}
\end{align*}
\vspace{-10pt}

\noindent where $P$ is the probability of occurrence and a scaling parameter $\alpha$ is added to the counts of each word. This scaling parameter $\alpha$ prevents single-count or one-off words in each cluster from having the highest PMI values. Given the scale of our dataset and the number of clusters within our dataset, we determine that a baseline count of 1 ($\alpha$ =1) for each word in the full dictionary in each cluster led to the best results~\cite{turney2001mining}.  We utilize this approach rather than cluster-normalized TF-IDF as in other works~\cite{grootendorst2022bertopic,hanley2022happenstance} because class TF-IDF is dependent on document classes being similar in length~\cite{aizawa2003information, grootendorst2022bertopic} and the number of articles within each of our clusters varies widely. PMI finds the distinct characteristics of individual clusters and is not dependent on how often words appear in other individual clusters, avoiding this issue.

\section{JS-Divergence\label{sec:js-divergence}}
Formally JS-Divergence between two distributions P and Q is calculated as follows:
\begin{align*}
    JS(P||Q)&=\frac{1}{2}KL(P||\frac{(P+Q)}{2})+\frac{1}{2}KL(Q||\frac{(P+Q)}{2}) \\
    KL(P||Q)&=\sum_{x}P(x)\log(\frac{P(x)}{Q(x)})
\end{align*}
For our purposes, given that every website does not address every topic, as recommended in other works, we add a small value $\epsilon=0.1$ to the counts of every website's topics before calculating each website's probability distribution.




\onecolumn
\section{Auto-Generated Summaries and Cluster Specificity\label{sec:cluster-spec}}

\scriptsize
\begin{tabularx}{\textwidth}{lsbb}
\toprule
Narr. &{Keywords} & Auto-Generated Summary & Random Sample Passage \\\midrule
1 & trudeau, motion, 151, 185, emergency, & \tiny{The Canadian Parliament voted Monday night to approve Prime Minister Justin Trudeau's motion to invoke the Emergencies Act by a vote of 185 for and 151 against.} & \tiny{On Monday night, Canada's parliament voted to confirm Prime Minister Trudeau's declaration of the Emergencies Act in response to the freedom protests that have swept across the nation for three weeks.} \\ \midrule

 2 & manchin, filibuster, schumer, sinema, senate & \tiny{Republicans and other critics immediately started to wonder: If Democrats extract what they want out of Manchin, couldn't even a small number of them promptly refuse to go along with the secondary assurances he's been promised? } &\tiny{[The pipeline Manchin was promised] would require passage of legislation that would overhaul the permitting process for energy infrastructure, according to The American Prospect, a liberal website. Apparently, progressives in the House are not keen on supporting a measure that could undercut the IRA s down payment on clean energy by accelerating approval for energy projects that could ramp up U.S. fossil fuel production and exports of natural gas, The American Prospect reported.} \\\midrule

 3 & agrawal, musk, parag, ceo, twitter& \tiny{Twitter CEO Parag Agrawal tweeted he was "excited" that Musk would join Twitter's board after it was revealed that Musk bought a 9.2 percent stake in the company, and in doing so became its largest shareholder.}  &  \tiny{When Musk's takeover of Twitter became official, Agrawal and Bret gave comments alongside the Tesla CEO.} \\ \midrule

 4 & capitol, committe, hearing, select, january&  \tiny{The House of Representatives committee investigating the Jan. 6, 2021, attack on the U.S. Capitol is planning to hold its next hearing on Sept. 28.} & \tiny{The U.S. House of Representatives select committee investigating the deadly Jan. 6, 2021, attack on the Capitol will conduct its next hearing on Oct. 13, the panel said in a statement on Thursday.}\\ \midrule

5 & smith, slap, black, oscar, rock& \tiny{Apparently, whites can't be outraged by Will Smith s slap without being racist. And never mind that plenty of blacks including Kareem Abdul-Jabbar were also outraged} & \tiny{I elaborated that Will Smith proved he believes violence is the way to handle disagreement. He makes blacks look bad  his slap reinforces the widely held stereotype that blacks are violent. He shamed AMPAS before the world. }\\\midrule

6  & extremist, maryland, walkby, virginia, protest& \tiny{The group that calls itself Ruth Sent Us announced its plans on a website to harass the justices. It said: Announcing Walkby Wednesday, May 11, 2022!At the homes of the six extremist justices, three in Virginia and three in Maryland.} & \tiny{RSU subsequently announced a WalkBy Wednesday protest on May 11, to be held in front of the homes of the six extremist justices} \\ \midrule

7 & cornyn, boo, convent, texas, gop &\tiny{U.S. Senator John Cornyn RTexas was loudly booed at the Republican Party of Texas Convention in Houston, where the state GOP adopted a resolution condemning the bipartisan gun control framework he has negotiated in the Senate.} & \tiny{Very loud boos for John Cornyn as he takes the stage at the TexasGOP convention. Cornyn has faced opposition within the party for working with Democrats on a gun package after the shooting in Uvalde.  }  \\\midrule

 8 & threat, truss, china, uk, britain& \tiny{Prime Minister Liz Truss is for the first time due to officially declare China a threat to the UK within days. The designation would be a formal update to former PM Boris Johnson s Integrated Review of Defense and Foreign Policy published in March 2021.} & \tiny{On October 11, The Guardian reported that the Liz Truss government is going to formally designate China a national "threat" to Britain in its upcoming strategic defense review. Under former Prime Minister Boris Johnson, China was named just a "systemic competitor.} \\ \midrule

9 &lithuania, vilniu, baltic, beijing, export & \tiny{China has called for a corporate boycott of the small Baltic nation. The move is in retaliation for Lithuania's decision to open a Taiwanese representative office in its capital of Vilnius in November 2021.} & \tiny{In a letter last month, the GermanBaltic Chamber of Commerce demanded that Lithuania come to a constructive solution with the communist nation, saying per Reuters: The basic business model of the companies is in question and some  will have no other choice than to shut down production in Lithuania.}\\\midrule

 10 & curfew, quebec, province, legault, montreal & \tiny{Quebec first imposed a COVID curfew on January 9, 2021, which was lifted on May 28. Quebec is the only province in Canada to have imposed a curfew during the pandemic} &  \tiny{Quebec is the first province to impose such a system on its citizens. It\'s also the only province in Canada that has a curfew in place.}\\ 

\bottomrule
\end{tabularx}\vspace{2pt}
\begin{center}
\centering
\hspace{15pt}
 Example of the auto-generated summaries and passages from a set of 10~random narrative clusters to illustrate the specificity and the precision of our approach.
\end{center}

\newpage
\twocolumn
\normalsize
\section{Paper Reviews}
\subsection{Summary of Paper}
This paper introduced a system to automatically track news narratives spread online. The paper analyzed news across over 1334 unreliable news sites and identified 52K narratives. Using the data, the authors examined news sites that amplify the narratives and showed how the information can be used to help with fact-checking.

\subsection{Scientific Contributions}
\begin{enumerate}
\item Creates a New Tool to Enable Future Science
\item Provides a Valuable Step Forward in an Established Field
\end{enumerate}

\subsection{Reasons for Acceptance}
\begin{enumerate}
    \item The paper provides a valuable step forward in an established field. The paper shows an end-to-end system to keep track of narratives of (fake) news over a large number of unreliable news websites. The authors collected months of data across more than 1334 news websites and identified over 52K narratives. The analysis provided new insights into the impact of the characteristics of the outlets on the efficacy of the propagation of narratives in terms of origination and amplification and the roles that social media sites (8kun and 4chan) play.
    \item The paper creates a new tool to enable future science. The narrative tracking system leverages a range of NLP and clustering tools. The ability to track different narratives can potentially help with identifying new misinformation for fact-checkers to audit. The authors plan to make the data available to researchers (and fact-checkers).
\end{enumerate}

\end{document}